\documentclass[11pt]{article}

\usepackage[T1]{fontenc}
\usepackage[utf8]{inputenc}
\usepackage{lmodern}
\usepackage{microtype}
\usepackage[margin=1in]{geometry}
\usepackage{graphicx}
\usepackage{booktabs}
\usepackage{array}
\usepackage{amsmath}
\usepackage{amssymb}
\usepackage{authblk}
\usepackage[font=small,labelfont=bf]{caption}
\usepackage{url}
\usepackage{flafter}
\usepackage{float}
\usepackage[section]{placeins}
\usepackage[hidelinks]{hyperref}
\hypersetup{
  pdftitle={Visualization and Characterization of Gene Group Distribution Patterns
            Across Genomes},
  pdfauthor={Yanying Wu},
  pdfsubject={Comparative genomics},
  pdfkeywords={gene families; gene groups; chromosome ideogram; interactive
               visualization; tandem duplication; gene spacing; comparative genomics}}

\graphicspath{{figs/}}

\newcommand{\Ce}{\textit{C.~elegans}}
\newcommand{\At}{\textit{Arabidopsis}}
\newcommand{\Dm}{\textit{D.~melanogaster}}
\newcommand{\wshort}{w_{\mathrm{short}}}

\title{\bfseries Visualization and Characterization of Gene Group\\
Distribution Patterns Across Genomes}

\author[ ]{Yanying Wu}
\affil[ ]{Centre for Neural Circuits \& Behaviour, University of Oxford, Oxford, United Kingdom}
\affil[ ]{\normalfont\small\href{mailto:yanying.wu@cncb.ox.ac.uk}{\texttt{yanying.wu@cncb.ox.ac.uk}}}
\date{\number\day~\ifcase\month\or January\or February\or March\or April\or May\or June\or July\or August\or September\or October\or November\or December\fi~\number\year}

\begin{document}
\maketitle

\begin{abstract}
\noindent
Genes of the same family often lie close together on a chromosome, yet how far apart family
members sit, and whether their spacing follows any pattern, has rarely been measured
consistently across genomes. To address this, we built an interactive web application that
places the members of any of the 1,396 FlyBase gene groups on the banded ideogram of the
fruit fly \textit{Drosophila melanogaster} and lists each member's neighbours.

While browsing it, we found strikingly symmetric gene sets in which consecutive members are
almost evenly spaced. In the RAB GTPases, for example, two 5.08~Mb spacings differ by only
634~bp. We also found that the distances between neighbouring members fall into two
populations: short-range (median $\approx$1.8~kb) and long-range (median $\approx$2.4~Mb).
Although an arm-matched random control suggests that the even spacing (when no minimum
distance between members is imposed) as well as the long-range population arise largely by
chance, the short-range population turns out to be a genuine clustering signal.

We then extended the analysis to human and six other model organisms (mouse, zebrafish,
\textit{Caenorhabditis elegans}, \textit{Saccharomyces cerevisiae}, \textit{Arabidopsis
thaliana} and maize), and every genome showed the same two populations. Interestingly, the
mixture curve of the two is nearly constant across chromosomes in some genomes (Type~I: fly,
maize, \textit{Arabidopsis}, \textit{C.~elegans}) but varies across chromosomes in others
(Type~II: human, mouse, zebrafish, yeast).

Thus, across eight eukaryotes, the distances between family members fall into two
populations. The scale of the short-range population follows the spacing of genes in each
genome, and the long-range population matches random placement in fly and is consistent with
it in the other genomes. Our results also show that visual patterns such as even spacing must be
tested against chance before they are interpreted. Overall, our tool and approach offer a
simple way to explore and compare gene clustering across species.
\end{abstract}

\section{Introduction}

Gene order along eukaryotic chromosomes is not random \cite{hurst}. Co-expressed and
housekeeping genes cluster into chromosomal domains in yeast, fly and human
\cite{cohen,spellman,boutanaev,lercher}, and many gene families have grown by tandem
duplication\footnote{Tandem duplication: duplication of a stretch of DNA into an adjacent
position, so that the new copy of a gene lies next to the old one \cite{zhang}.} and still
sit together where they arose \cite{lespinet,shoja,rizzon}. Other families are dispersed,
having been spread by segmental or whole-genome duplication\footnote{Whole-genome
duplication: duplication of an entire genome, after which most duplicate genes are
lost and the survivors, called ohnologs, are scattered across chromosomes
\cite{wolfe,jaillon}.} \cite{wolfe,jaillon,lynch}, by transposition, or by chromosomal
rearrangement \cite{bhutkar}. A family can therefore be clustered, dispersed, or both. Yet
how far apart family members sit, and whether their spacing follows any pattern, has rarely
been measured consistently across genomes.

A simple way to see where genes lie is to draw them on an ideogram\footnote{Ideogram: a
schematic drawing of the chromosomes of a genome, to scale, usually with their cytogenetic
bands.} \cite{ideogram}. We built an interactive web application on which the members of
any of the 1,396 FlyBase gene groups \cite{flybase} can be displayed on the banded ideogram
of the fruit fly \textit{Drosophila melanogaster} (\S\ref{sec:app}). Browsing it, two impressions recur
(\S\ref{sec:browse}). Some groups contain sets of members that appear evenly spaced, with a
middle gene almost equidistant from its two neighbours. And some members sit in tight
clusters, within a few kilobases of each other, while others are megabases apart, often on
different arms.

This paper turns those impressions into measurements. For every gene family we take each
pair of consecutive members on the same chromosome or arm, a pair of \emph{neighbour genes},
and measure the distance between them. In fly we first ask how common evenly spaced,
symmetric sets are (\S\ref{sec:symmetric}), and what shape the neighbour-distance
distribution has (\S\ref{sec:twopop}). To test whether these patterns exceed chance, we used
a random control that replaces each group with the same number of random genes from the same
chromosome arms (\S\ref{sec:randctl}). This control produced about 85\% as many symmetric
three-gene sets when no minimum distance between members is imposed, so even spacing arises
largely by chance (\S\ref{sec:symcontrol}). The control also
reproduced the long-range population but not the short-range one. The short-range
population is therefore a genuine clustering signal, most likely from tandem duplication
(\S\ref{sec:control}).

We then extend the analysis to human and six other model organisms (\S\ref{sec:species}),
the same eight genomes whose chromosomes we compared for functional element composition in a
companion study \cite{wu2026}, and ask whether the mixture of the two populations is the same on every chromosome of a
genome (\S\ref{sec:types}), and what sets the short-range distance scale: genome size, gene
size, or the spacing of genes (\S\ref{sec:h3}). The short-range distance runs from 1.8~kb in
yeast and fly to 69~kb in maize, and it follows the spacing of genes rather than genome size or
gene length (\S\ref{sec:h3}). Finally, we show that differences in the weights of the two
populations between genomes come mostly from group size, and then from clade
(\S\ref{sec:weights}).

\section{Data and methods}
\label{sec:methods}

\subsection{Genomes, genes and gene families}

Eight assemblies were used, the same as in the companion study of functional element
composition \cite{wu2026}, each in its own coordinates with no liftover:
\Dm{} dm6 (release~6 of the Berkeley \textit{Drosophila} Genome Project, BDGP6)
\cite{adams,hoskins}, human (\textit{Homo sapiens}) GRCh38 (Genome Reference Consortium
human build~38) \cite{lander,schneider}, mouse (\textit{Mus musculus}) GRCm39
\cite{mouse}, zebrafish (\textit{Danio rerio}) GRCz11 \cite{howe}, the budding yeast
\textit{Saccharomyces cerevisiae} R64-1-1 \cite{goffeau}, the nematode
\textit{Caenorhabditis elegans} WBcel235 \cite{celegans}, \textit{Arabidopsis thaliana}
TAIR10 (The \textit{Arabidopsis} Information Resource release~10) \cite{agi,tair} and
maize (\textit{Zea mays}) Zm-B73-REFERENCE-NAM-5.0, the B73 assembly of the maize nested
association mapping (NAM) founder lines \cite{schnable,hufford}. Chromosome names follow FlyBase/Ensembl
style without a \texttt{chr} prefix. All coordinates were converted to 0-based half-open
intervals at load; GFF3 (General Feature Format version~3), FlyBase and BioMart coordinates
are 1-based inclusive and were shifted accordingly. Table~\ref{tab:data} lists the sources.

\paragraph{Gene coordinates.} Fly gene coordinates are the FlyBase dm6 gene list
\cite{flybase}. Human, mouse, zebrafish, yeast and \Ce{} coordinates come from Ensembl
BioMart \cite{ensembl}; \At{} and maize coordinates were parsed from Ensembl Plants GFF3
files \cite{ensemblgenomes}, because the Ensembl Plants BioMart service failed every query
at the time of download.

\paragraph{Gene families.} A \emph{gene group} is any set of genes that a resource assigns to
one family. Fly uses the 1,396 curated FlyBase gene groups (release FB2026\_01) that have at
least one placed member \cite{flybase}, and human the 2,153 curated gene groups of the HUGO (Human Genome
Organisation) Gene Nomenclature Committee (HGNC) \cite{hgnc}. No downloadable curated table of the same kind exists for the other six
genomes, so for them a group is an InterPro entry of type \emph{Family} \cite{interpro},
assigned to genes through Ensembl BioMart or, for \At{} and \Ce, through the UniProt
reference proteomes \cite{uniprot}. To test whether results depend on the group source we
also built (i)~InterPro \emph{Family} groups for human, on the same coordinates and arms as
the HGNC groups; (ii)~groups from WormBase gene-class names (\texttt{nhr}, \texttt{srh},
\texttt{col}, \ldots) for \Ce{} \cite{wormbase}; and (iii)~a looser definition,
InterPro \emph{Family}~+~\emph{Domain}, for every InterPro genome. HGNC groups include a few
large biotype collections (antisense RNAs, long non-coding RNAs, microRNAs, pseudogene
classes) that are not families; they are kept in the primary analysis and removed in a
robustness check (\S\ref{sec:robust}).

\begin{table}[!htbp]
\centering
\small
\caption{Genomes and gene-group sources. Genome size is the assembled length used
throughout. Groups: number of groups with at least one placed member. The unit is the
stretch of chromosome within which neighbour genes are paired.}
\label{tab:data}
\begin{tabular}{@{}llrlrl@{}}
\toprule
Species & Assembly & Size (Mb) & Group source & Groups & Unit \\
\midrule
\textit{S.~cerevisiae} & R64-1-1 & 12.1 & InterPro Family & 3,475 & 16 chromosomes \\
\Ce & WBcel235 & 100.3 & InterPro Family & 4,656 & 6 chromosomes \\
\At & TAIR10 & 135.0 & InterPro Family & 4,999 & 5 chromosomes \\
\Dm & dm6 & 143.7 & FlyBase gene groups & 1,396 & 6 arms \\
zebrafish & GRCz11 & 1,373 & InterPro Family & 7,513 & 25 chromosomes \\
maize & Zm-B73 NAM-5.0 & 2,178 & InterPro Family & 4,623 & 10 chromosomes \\
mouse & GRCm39 & 2,723 & InterPro Family & 8,648 & 21 chromosomes \\
human & GRCh38 & 3,100 & HGNC gene groups & 2,153 & 48 p/q arms \\
\bottomrule
\end{tabular}
\end{table}

\paragraph{Units.} Neighbour genes are paired within a \emph{unit}. In fly the unit is the
chromosome arm (\texttt{X}, \texttt{2L}, \texttt{2R}, \texttt{3L}, \texttt{3R},
\texttt{4}); in human it is the p or q arm, read from the leading letter of each gene's
cytogenetic band. Splitting at the centromere keeps a pair from spanning megabases of
centromeric heterochromatin. In the other six genomes the unit is the whole chromosome:
mouse chromosomes are acrocentric, with no annotated genes on the short arm \cite{mouse}; yeast
centromeres are point centromeres of about 125~bp \cite{fitzgerald}; \Ce{} chromosomes are holocentric
\cite{maddox}; and the zebrafish, \At{} and maize annotations used here carry no centromere
positions. A pair in these genomes may therefore span a centromere.

\subsection{Ideograms and the interactive web application}
\label{sec:appmethods}

The fly ideogram is drawn from the dm6 cytogenetic (Bridges) band map \cite{bridges}
distributed with \texttt{ideogram.js} \cite{ideogram}, with rounded telomeres and hourglass
centromeres; arms \texttt{2L}/\texttt{2R} and \texttt{3L}/\texttt{3R} are joined at their
centromeres. No band map was used for the other seven genomes, whose chromosomes are drawn as
plain bars scaled to assembled length. In the whole-genome views every located gene is drawn
at its position and coloured by gene class, using the gene classification and colour palette
of \cite{wu2026}.

The web application is a \texttt{Panel} \cite{panel} application rendered with
\texttt{matplotlib} \cite{matplotlib}, served from notebook \texttt{ideograms/03} of the project repository.
Its controls are listed in \S\ref{sec:app}. All of its group-level computations ---
members, neighbour pairs and symmetric sets --- call the same library functions
(\texttt{fly\_genome.io}) that the analyses of this paper use, so a set seen in the
application is a set counted in the statistics.

\subsection{Symmetric gene sets}
\label{sec:symdef}

Symmetry is judged on midpoint-to-midpoint spacings, so that a gene's own length does not
enter. A \emph{symmetric three-gene set} is three consecutive members of one group on one
arm --- left, middle and right --- whose two flanking midpoint spacings $\delta_\ell$ and
$\delta_r$ satisfy
\begin{equation}
  \frac{|\delta_\ell - \delta_r|}{\max(\delta_\ell, \delta_r)} \le \tau,
\end{equation}
for a symmetry tolerance $\tau$. A \emph{symmetric four-gene set} is four consecutive members
$a, b, c, d$ on one chromosome, either all on one arm or $a, b$ on one arm and $c, d$ on the
other, whose two outer spacings ($a$--$b$ and $c$--$d$) satisfy the same rule. Optionally
every compared spacing must be at least a minimum distance $d_{\min}$. The left-hand side is
the set's \emph{asymmetry}; 0 is perfect symmetry. The application's defaults are
$\tau = 15\%$ and $d_{\min} = 10$~kb.

\subsection{Neighbour genes and neighbour distances}
\label{sec:neighbourdef}

For each group, the members on each unit are ordered by midpoint. Two members are
\emph{neighbour genes} if they lie on the same unit and no other member of the same group
lies between them; each such consecutive pair is a \emph{neighbour pair}. A group with
$n$ members on a unit therefore has $n-1$ neighbour pairs there, and members alone on their
unit have none. The \emph{neighbour distance} is the edge-to-edge distance
\begin{equation}
  d = s_b - e_a,
\end{equation}
where $e_a$ is the end of the first gene and $s_b$ the start of the second. A distance is
negative when the two genes overlap. As a check on unit assignment, no distance may exceed
the span of its unit. Distances run over five or more orders of magnitude, so they are
displayed as $\operatorname{sign}(d)\,\log_{10}(|d|+1)$, which shows overlaps and megabase
distances on one axis. (Axis labels and some panel titles in the figures read ``gap'' or ``intergenic gap''; it is
the same quantity.)

\subsection{Distribution fits}

Four positive-support families were fitted to the positive neighbour distances by maximum
likelihood with the location parameter fixed at zero: lognormal, gamma, Weibull and
exponential. The exponential is the null of interest, since it is the distance distribution
of points placed at random (a Poisson process) along a line. Models were ranked by the Akaike
information criterion (AIC) \cite{akaike}, and the Kolmogorov--Smirnov (KS) statistic $D$,
the largest difference between the fitted and empirical cumulative distribution functions,
was used as a measure of shape mismatch.

Because the distances turned out to be bimodal, we also fitted Gaussian mixtures to
$\log_{10} d$ with $k = 1,\dots,4$ components, compared by the Bayesian information
criterion (BIC) \cite{schwarz}. A Gaussian on $\log_{10} d$ is a lognormal in base pairs,
so $k = 2$ is a mixture of a short-range and a long-range lognormal. We report the
median (the back-transformed mean) and weight of each component, and denote the weight of
the short-range component $\wshort$. BIC usually kept improving up to $k = 4$, but almost
all of the gain came from the step $k = 1 \to 2$ (Table~\ref{tab:summary}); we report
$k = 2$ throughout because it is interpretable and comparable between genomes.
Fits used \texttt{scipy} \cite{scipy} and \texttt{scikit-learn} \cite{sklearn}
(five initialisations, fixed seed).

\subsection{Arm-matched random control}
\label{sec:ctlmethods}

To separate the effect of gene-family biology from that of chromosome geometry, each fly
gene group was replaced by a control set of the same size and arm composition: for every
arm, the same number of genes the group has on that arm were drawn uniformly at random,
without replacement, from all dm6 genes on that arm. The control therefore has the same
number of sets, set sizes and neighbour pairs as the real groups, but no positional
structure. The full analysis was repeated on the control (seed~0), and the two
neighbour-distance distributions were compared by a two-sample KS test. Symmetric sets were
compared over 20 independent control draws.

\subsection{Per-unit fits and genome types}
\label{sec:typemethods}

The same $k = 2$ mixture was fitted separately on
every unit with at least 50 positive distances (25 for yeast \emph{Family} groups). A forced
two-component fit always returns two components, so a unit was called bimodal only if the
component means were at least 0.5~dex (orders of magnitude, i.e.\ $\log_{10}$ units) apart and neither weight was below 0.10. Each component
median was then related to the gene-bearing span of its unit by Spearman's rank correlation
$\rho$ and by a log--log regression. To summarise how consistent the per-unit mixtures of a
genome are, we computed, over its bimodal units, the standard deviation (SD) of
$\log_{10}$ of the short-range median, the SD of $\log_{10}$ of the long-range median, and
the SD of $\wshort$. A genome is called \emph{Type~I} (consistent) when all three are small
--- at most 0.25~dex, 0.15~dex and 0.085 respectively --- and \emph{Type~II} (variable)
otherwise. The thresholds are descriptive; they sit in the clear break between the two
groups of genomes (Table~\ref{tab:types}).

\subsection{Gene size and gene spacing}

Two genome-wide quantities were measured from the whole gene pool of each genome, not from
the group members alone: the median gene span (start to end, introns included) and the
median distance between consecutive genes, the \emph{next-gene distance}, taken over all
placed genes sorted by position on each unit. Overlapping or nested gene pairs give
non-positive distances and were excluded. In \Ce{} the 15,363 21U-RNA (PIWI-interacting RNA, piRNA) loci, 21~bp
each and packed into two clusters on chromosome \texttt{IV}, were excluded from the pool; no
grouped gene can be a neighbour to them.

\subsection{Size-matched component weights}

A group with $n$ members on one unit contributes $n-1$ neighbour distances, of which only a
few can be short-range, so large groups lower $\wshort$ by construction. Group tables differ
widely in their size distributions (HGNC groups have on average 16.3 placed members, InterPro
families 4.9), so raw weights are not comparable between genomes. For a fair comparison,
every genome was restricted to groups of 2--10 placed members (and, as checks, 2--5, 2--20
and 3--10), and the mixture was refitted.

\section{Visualizing gene groups on the fly ideogram: an interactive web application}
\label{sec:app}

\begin{figure}[!htbp]
\centering
\includegraphics[height=3.7in]{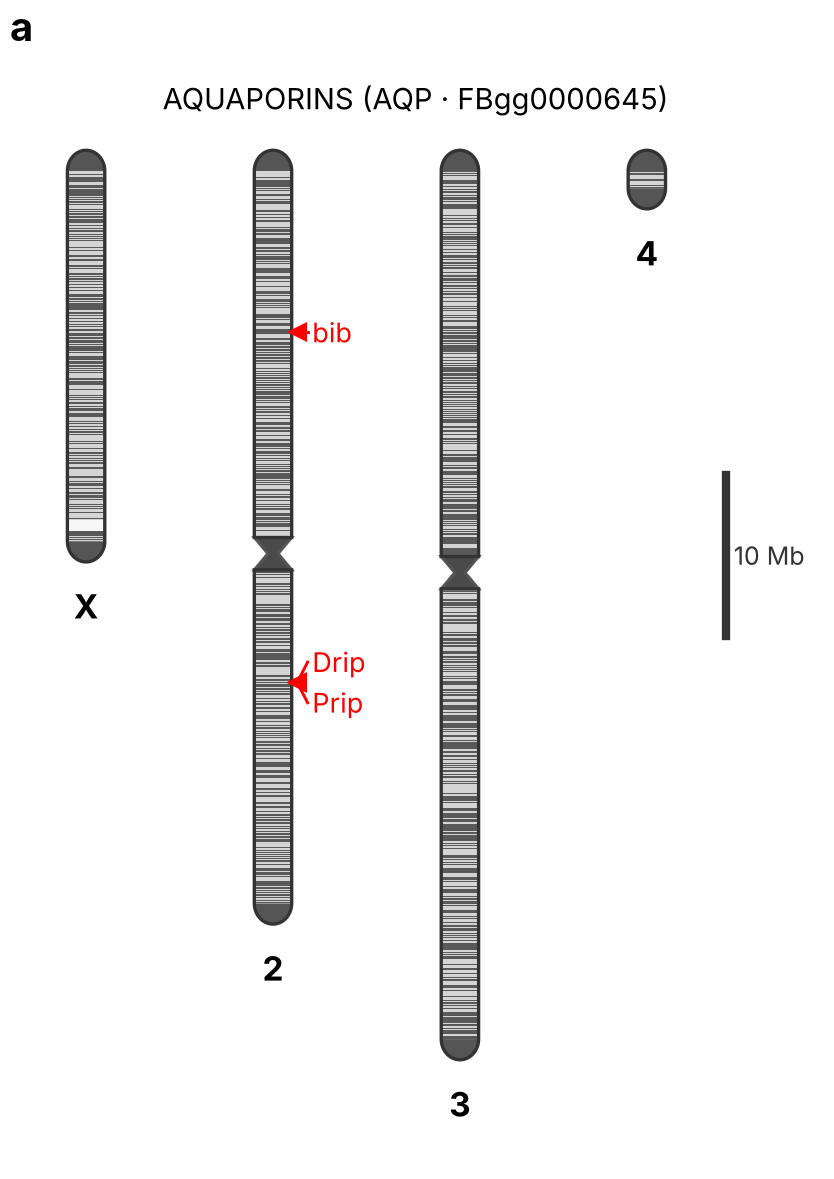}\hfill
\includegraphics[height=3.7in]{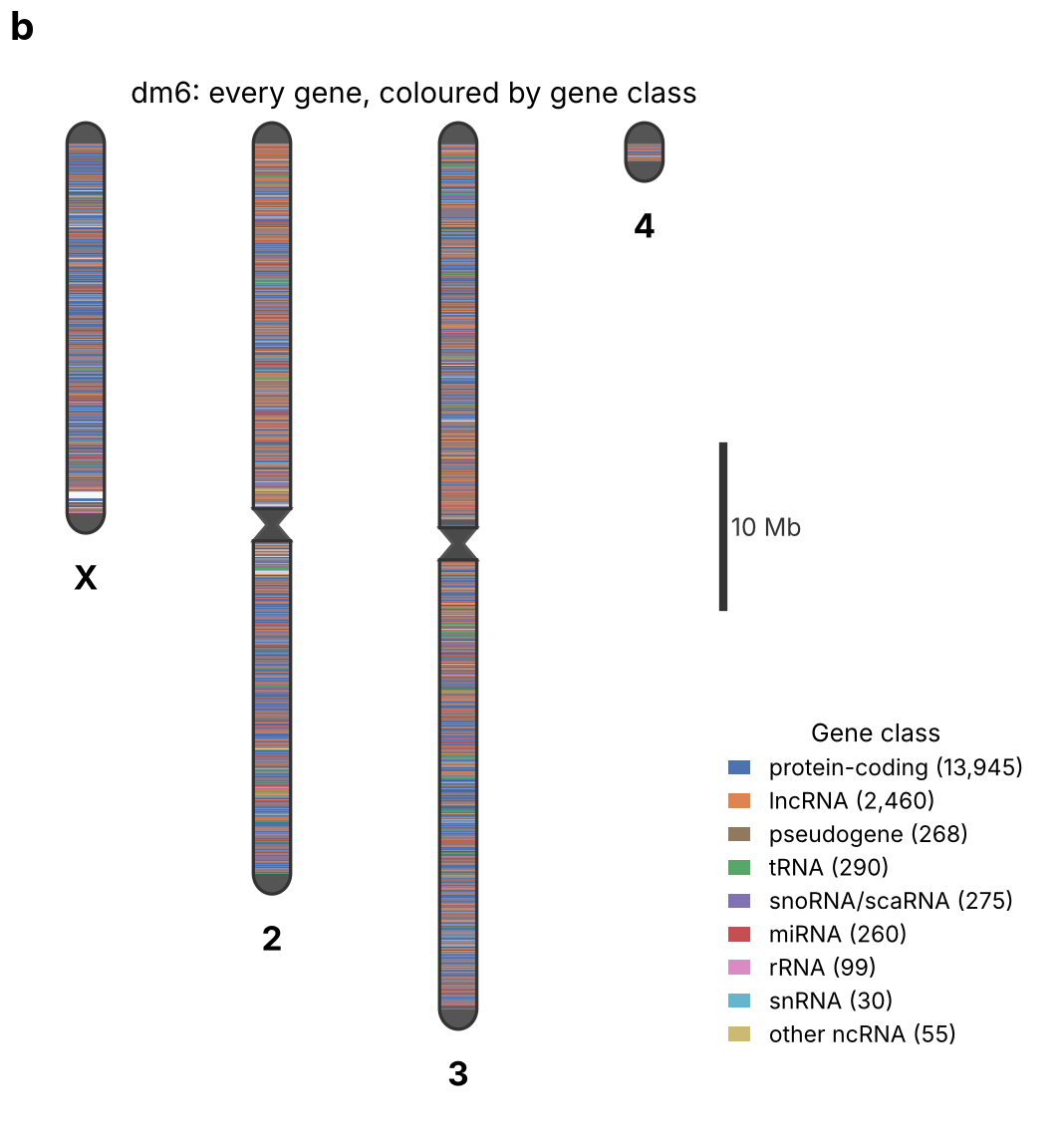}
\caption{The \Dm{} genome (dm6) on an ideogram. (a)~Banded ideogram with the members of
one FlyBase gene group, the aquaporins (AQP), marked; two members sit side by side on
\texttt{2R} and the third on \texttt{2L}. (b)~Every located fly gene drawn at its own
position, coloured by gene class: protein-coding, long non-coding RNA (lncRNA), pseudogene,
and the small non-coding RNA (ncRNA) classes, i.e.\ transfer (tRNA), small nucleolar and small
Cajal body-specific (snoRNA/scaRNA), micro (miRNA), ribosomal (rRNA) and small nuclear RNA
(snRNA); classification as in \cite{wu2026}. Arms \texttt{2L}/\texttt{2R} and \texttt{3L}/\texttt{3R} are joined at the
centromere.}
\label{fig:ideogram}
\end{figure}

Figure~\ref{fig:ideogram} shows the fly genome on its banded ideogram, the frame on which the
rest of the paper builds. On the left, one gene group, the aquaporins, is marked:
\textit{bib} on \texttt{2L} and the adjacent pair \textit{Drip}, \textit{Prip} on
\texttt{2R}, about 7~kb apart. On the right, every located fly gene is drawn at its position
and coloured by class. Genes cover the euchromatic arms densely and thin out towards the
pericentromeric heterochromatin.

Gene content is broadly similar across the five large fly arms (Appendix
Fig.~\ref{fig:armstats}): they are 23.5--32.1~Mb long, carry 2,639--4,147 genes at
112--147 genes/Mb, and are 63--72\% genic by base pairs. Chromosome~\texttt{4} has the
fewest genes (110) but the highest genic coverage (79\%), because its genes are long, and
\texttt{Y} is gene-poor (31 genes/Mb, 30\% genic). Gene length is close to lognormal on
every arm (median 1.5--2~kb); a minor peak near 100~bp is formed by the short ncRNA classes
(tRNA, miRNA, snoRNA, rRNA, snRNA; Appendix Fig.~\ref{fig:biotype}).

\begin{figure}[!htbp]
\centering
\includegraphics[width=\textwidth]{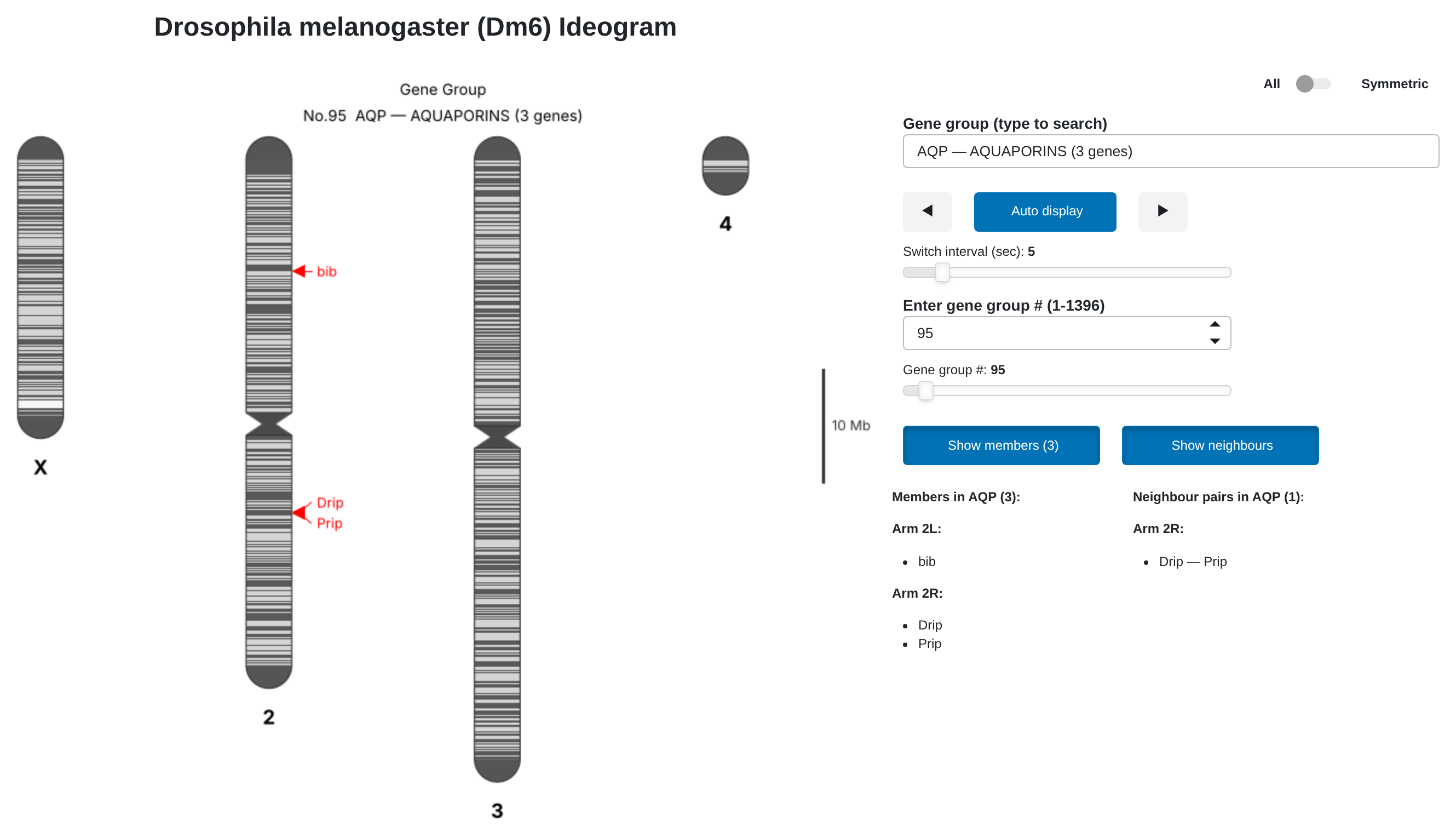}
\caption{The interactive gene-group ideogram, in \emph{All} mode. The aquaporin group
(No.~95, AQP, three genes) is selected: its members are marked on the banded dm6 ideogram
(left), and the member list and neighbour-pair list are open (right). The controls are, from
the top: the All/Symmetric switch, the searchable gene-group box, previous/next buttons and
automatic display, the display interval, the group-number box and slider, and the
\emph{Show members} and \emph{Show neighbours} buttons.}
\label{fig:app}
\end{figure}

To see where the members of a gene family lie, and how they are spread along the genome, we
built an interactive web application on the banded fly ideogram (Fig.~\ref{fig:app}). Any of
the 1,396 FlyBase gene groups can be chosen from a searchable box, and its members are marked
on the ideogram by arrows with gene names; labels are spaced automatically so that they never
overlap, and groups with more than 30 members are shown by markers alone. The group can also
be chosen by its number, with a slider, with previous/next buttons, or by an automatic
display that steps through the groups at a chosen interval (3--20~s). Two toggle buttons list
the group's members arm by arm, in ideogram order, and its neighbour pairs --- the
consecutive members on each arm (\S\ref{sec:neighbourdef}). The application runs locally
with a single command and needs only a web browser.

\begin{figure}[!htbp]
\centering
\includegraphics[width=\textwidth]{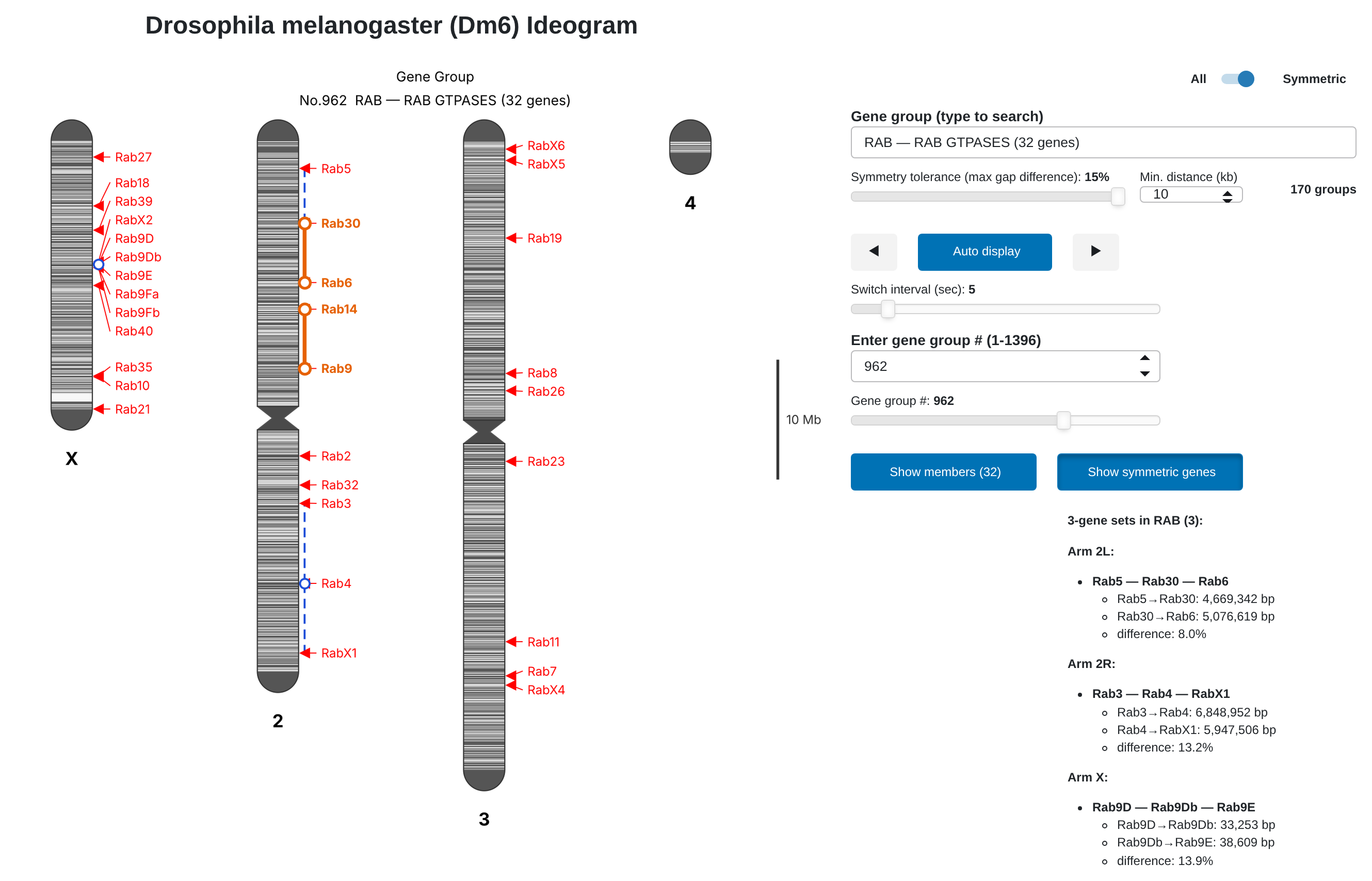}
\caption{The application in \emph{Symmetric} mode, at the default tolerance (15\%) and
minimum distance (10~kb). The RAB GTPase group (No.~962, 32 genes) is selected; gene names,
which the application hides for groups of more than 30 members, are shown here. Blue dashed connectors and rings mark its symmetric sets on \texttt{X},
\texttt{2L} and \texttt{2R}; the list on the right gives each three-gene set with its two
midpoint spacings and their difference. The four-gene set Rab30--Rab6~$\cdot$~Rab14--Rab9 on
\texttt{2L}, the most symmetric set in the fly genome (Table~\ref{tab:topsym}), is highlighted
in orange (bold names, solid connectors and rings) on the upper arm of chromosome~2.}
\label{fig:appsym}
\end{figure}

A switch above the search box changes the application to \emph{Symmetric} mode
(Fig.~\ref{fig:appsym}). The search list then narrows to groups that contain at least one
symmetric three- or four-gene set (\S\ref{sec:symdef}); a slider sets the symmetry tolerance
(0--15\%), a box sets the minimum distance (10--100~kb), and a live counter shows how many
groups qualify (170 at the defaults). The ideogram draws dashed connectors for every
symmetric set of the selected group --- for a three-gene set, from the middle gene to its two
flankers, with the middle gene ringed; for a four-gene set, across each of the two matched
pairs --- and the neighbour button becomes \emph{Show symmetric genes}, which lists each set
with its two spacings and their percentage difference. Group numbers stay those of the full
list, so a group can be found again in either mode.

\section{Two patterns seen by browsing}
\label{sec:browse}

Browsing the application group by group gives two recurring impressions: some members look
evenly spaced, and members tend to be either very close together or megabases apart. We
describe each in turn, and test both against chance in \S\ref{sec:randctl}.

\subsection{Symmetric gene sets}
\label{sec:symmetric}

\begin{figure}[!htbp]
\centering
\includegraphics[width=0.96\textwidth]{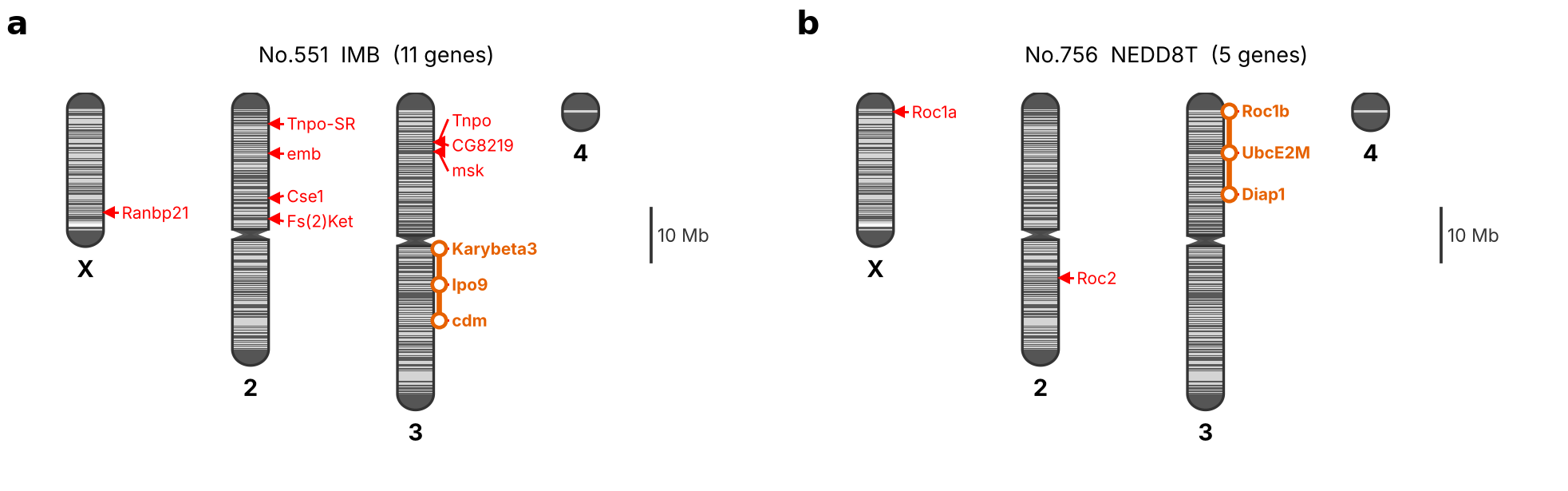}
\captionsetup{font=footnotesize}
\caption{Two of the six most symmetric sets (Table~\ref{tab:topsym}): the beta importins and
the NEDD8 transferases, shown by the application in Symmetric mode ($\tau = 15\%$,
$d_{\min} = 10$~kb). In each panel the set of Table~\ref{tab:topsym} is highlighted in orange
(bold names, solid connectors and rings), as in Fig.~\ref{fig:appsym}. Spacings $\delta_1$,
$\delta_2$ are midpoint to midpoint. (a)~Beta importins (No.~551, 11 genes): three-gene set
Karybeta3--Ipo9--cdm on \texttt{3R}, $\delta_1 = 6{,}762{,}432$~bp,
$\delta_2 = 6{,}767{,}045$~bp, asymmetry 0.068\%; Ipo9 lies within 2.3~kb of the midpoint of
the 13.5~Mb interval. (b)~NEDD8 transferases (No.~756, 5 genes): three-gene set
Roc1b--UbcE2M--Diap1 on \texttt{3L}, $\delta_1 = 7{,}838{,}684$~bp,
$\delta_2 = 7{,}847{,}805$~bp, 0.116\%. The other four sets of Table~\ref{tab:topsym} are not
drawn. The RAB set (Rab30--Rab6~$\cdot$~Rab14--Rab9) is already shown in
Fig.~\ref{fig:appsym}. In the SLP-C-P, HOXL and CAT sets the orange members lie too close
together on a whole-genome ideogram for their spacings to be seen: the SLP-C-P and CAT sets span 4.8
and 5.1~Mb and bcd--Dfd--Scr 79~kb (against 13.5 and 15.7~Mb in a and b), so their rings and
labels run into one another.
The ideograms were redrawn with the application's own plotting functions.}
\label{fig:topsym}
\end{figure}

Browsing the application in Symmetric mode shows how common evenly spaced members are. At
the defaults ($\tau = 15\%$, $d_{\min} = 10$~kb), 170 of the 1,396 fly groups contain at
least one symmetric set, 396 sets in all; without the minimum distance, 240 groups contain
696 sets. Some are very tight. Table~\ref{tab:topsym} lists the six most symmetric sets in the
fly genome at the application's defaults, each from a different group. The tightest, the
RAB set, is shown in Fig.~\ref{fig:appsym}; Fig.~\ref{fig:topsym} shows two more, the beta
importins and the NEDD8 transferases, whose members are far enough apart on the whole-genome
ideogram for the two equal spacings to be seen. The other three are not drawn, because at this
scale their highlighted members lie too close together for the spacings to be judged by eye:
the sets of the spliceosome complex~P (SLP-C-P) and the cationic amino acid transporters
(CAT) span 4.8 and 5.1~Mb, and bcd--Dfd--Scr of the Hox-like homeobox transcription factors
(HOXL) only 79~kb, against 13.5 and 15.7~Mb for the two sets shown, so their rings and labels
run into one another and, for SLP-C-P, into those of the group's other 82 genes. In the RAB GTPases, Rab30--Rab6 and
Rab14--Rab9 on \texttt{2L} are 5,076,619 and 5,077,253~bp apart, a difference of 634~bp, or
0.012\%. The beta importins Karybeta3, Ipo9 and cdm sit on \texttt{3R} with Ipo9 within
2.3~kb of the midpoint of a 13.5~Mb interval.

\begin{table}[!htbp]
\centering
\small
\caption{The six most symmetric sets in the fly genome, one per group, at the application's
defaults ($\tau = 15\%$, $d_{\min} = 10$~kb). Spacings are midpoint to midpoint. Asymmetry:
$|\delta_1-\delta_2|/\max(\delta_1,\delta_2)$. FlyBase group abbreviations: RAB, RAB GTPases;
SLP-C-P, spliceosome complex~P; IMB, beta importins; NEDD8T, NEDD8 transferases; HOXL, Hox-like
homeobox transcription factors; CAT, cationic amino acid transporters.}
\label{tab:topsym}
\setlength{\tabcolsep}{4pt}
\begin{tabular}{@{}llllrrr@{}}
\toprule
Group (No.) & Arm & Kind & Set & $\delta_1$ (bp) & $\delta_2$ (bp) & Asym. \\
\midrule
RAB (962) & \texttt{2L} & four & Rab30--Rab6 $\cdot$ Rab14--Rab9 & 5,076,619 & 5,077,253 & 0.012\% \\
SLP-C-P (1145) & \texttt{2L} & four & snRNA:U5:23D--CG15432 $\cdot$ Phf5a--SmE & 1,393,887 & 1,394,691 & 0.058\% \\
IMB (551) & \texttt{3R} & three & Karybeta3--Ipo9--cdm & 6,762,432 & 6,767,045 & 0.068\% \\
NEDD8T (756) & \texttt{3L} & three & Roc1b--UbcE2M--Diap1 & 7,838,684 & 7,847,805 & 0.116\% \\
HOXL (523) & \texttt{3R} & three & bcd--Dfd--Scr & 39,478 & 39,419 & 0.151\% \\
CAT (192) & \texttt{3L} & three & CG7255--CG5535--CG13248 & 2,546,984 & 2,554,126 & 0.280\% \\
\bottomrule
\end{tabular}
\end{table}

Five of the six tightest sets span megabases. The exception is bcd--Dfd--Scr, in the
Antennapedia complex on \texttt{3R}: three Hox-like genes about 39~kb apart in a cluster that
arose by tandem duplication. The same group holds a second set, Ubx--abd-A--Abd-B, in the
bithorax complex (6.1\% asymmetry).

\subsection{Neighbour distances form two populations}
\label{sec:twopop}

\begin{figure}[!htbp]
\centering
\includegraphics[width=0.49\textwidth]{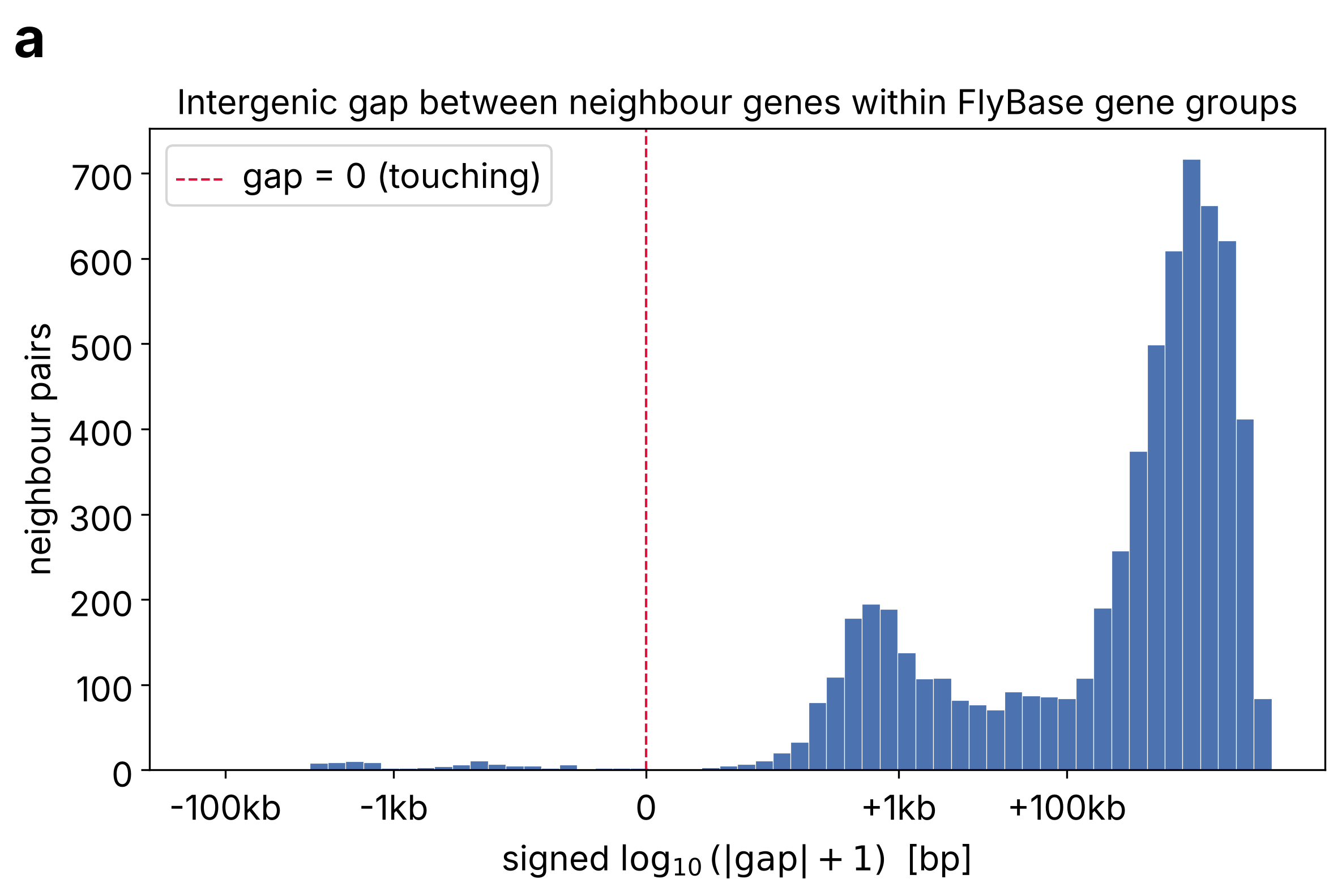}\hfill
\includegraphics[width=0.49\textwidth]{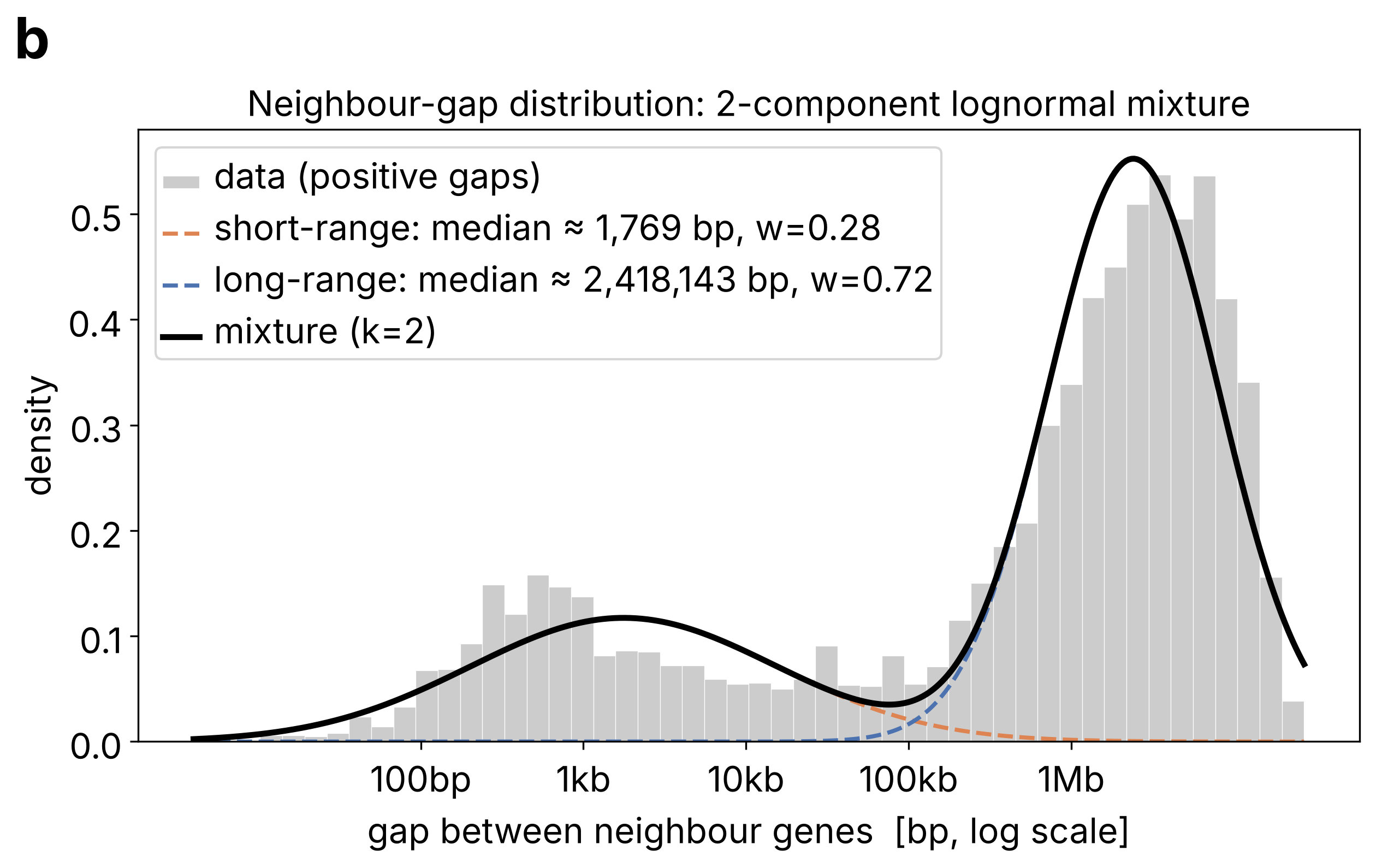}
\caption{Distances between neighbour genes of FlyBase gene groups (6,393 pairs).
(a)~Histogram of $\operatorname{sign}(d)\log_{10}(|d|+1)$; negative values are overlapping
pairs. (b)~Two-component lognormal mixture fitted to the positive distances: short-range
component, median 1.77~kb, weight 0.28; long-range component, median 2.42~Mb, weight 0.72.}
\label{fig:fly}
\end{figure}

The second impression, tight clusters next to members spread over megabases, concerns the
distance between neighbour genes (\S\ref{sec:neighbourdef}). The 1,396 fly gene groups yield
6,393 neighbour pairs. Of these, 99 (1.5\%) overlap. The
positive neighbour distances are clearly bimodal (Fig.~\ref{fig:fly}a): a minor population
of about 1--2~kb and a dominant one in the megabase range. No single distribution fits. Gamma is
the best by AIC, but with $D = 0.096$, and lognormal ($D = 0.19$) and exponential
($D = 0.24$) are worse; every unimodal curve misses the short-range peak. A two-component
lognormal mixture lowers the BIC by 4,751 relative to one component and separates a
short-range population with median 1.77~kb and weight 0.28 from a long-range population
with median 2.42~Mb and weight 0.72 (Fig.~\ref{fig:fly}b).

\section{A random control separates biology from geometry}
\label{sec:randctl}

Both patterns raise the same question: how much of each is due to gene families, and how
much to the simple fact of choosing a few genes on a fly arm? To answer it, each fly gene
group was replaced by the same number of genes drawn at random from the same chromosome
arms (\S\ref{sec:ctlmethods}), and the analysis was repeated on these controls.

\subsection{Symmetric sets are mostly chance}
\label{sec:symcontrol}

\begin{figure}[!htbp]
\centering
\includegraphics[width=\textwidth]{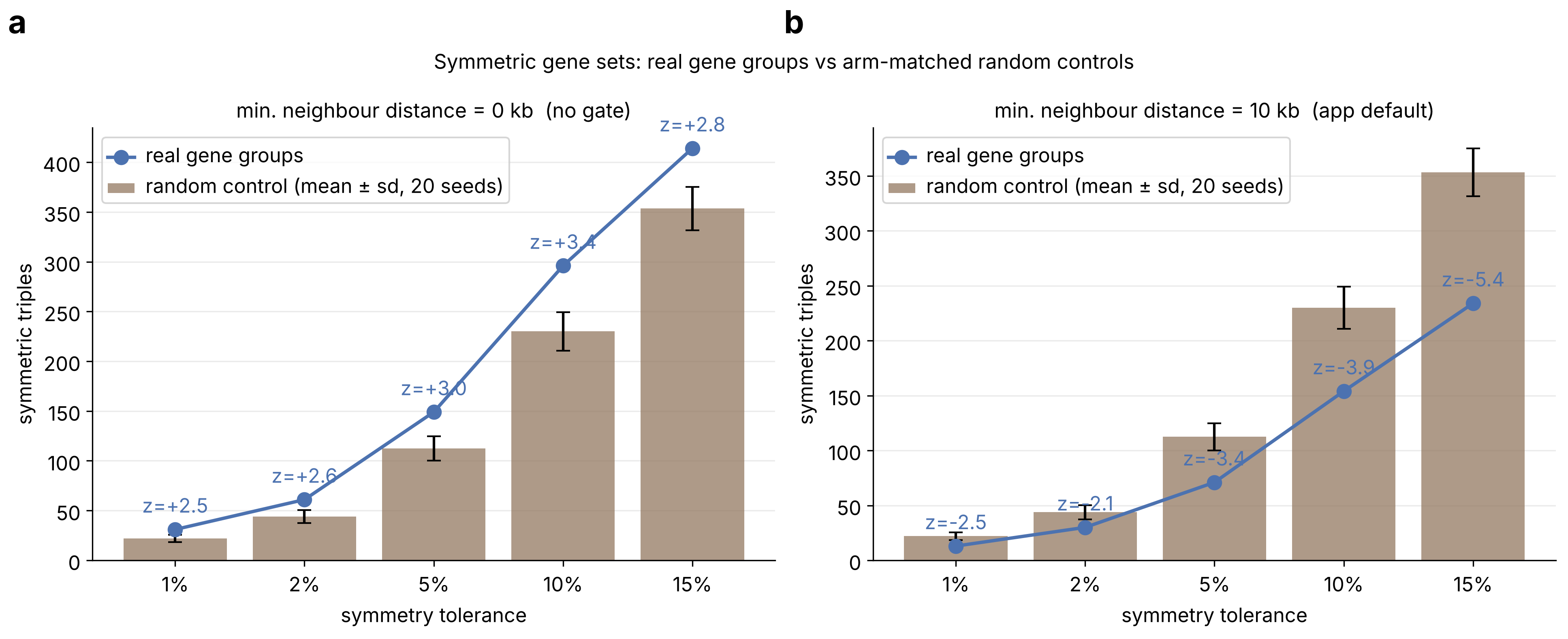}
\caption{Symmetric three-gene sets in real fly gene groups (line) and in 20 arm-matched
random controls (bars, mean $\pm$ SD), at symmetry tolerances of 1--15\%. (a)~No minimum
neighbour distance: real groups are modestly enriched ($z = +2.5$ to $+3.4$). (b)~Minimum
neighbour distance of 10~kb: real groups are depleted ($z = -2.1$ to $-5.4$), because the
gate removes the tandem-scale sets that only real groups contain.}
\label{fig:symmetric}
\end{figure}

Are symmetric sets more common than chance would produce (Fig.~\ref{fig:symmetric})?
Counting three-gene sets only, at $\tau = 15\%$ with no minimum distance, 186 fly groups contain at least one symmetric set, 414 sets in all (the other 282 of the 696
sets counted in \S\ref{sec:symmetric} are four-gene sets). The
control yields 353 $\pm$ 22 sets (mean $\pm$ SD, 20 draws), so chance accounts for about
85\% of them. The real excess is consistent but small: $z = +2.5$ to $+3.4$ at every
tolerance from 1\% to 15\%.

With a minimum distance of 10~kb, the default of the application, the sign reverses: 234 real
three-gene sets against 353 $\pm$ 22 in the control ($z = -5.4$; Fig.~\ref{fig:symmetric}b).
The reason is mechanical. The gate removes 43\% of the real sets, those with a neighbour
closer than 10~kb, and none of the control sets, since random genes are almost never that
close. For the same reason the symmetric sets that pass a 10~kb gate and a 5\% tolerance are
dominated by widely spaced members: of 131 such sets (71 three-gene and 60 four-gene sets),
74\% have a first spacing of 1--10~Mb (Appendix Fig.~\ref{fig:symfirst}). A tight relative
tolerance is easy to meet when genes are megabases apart and hard when they are a few
kilobases apart.

A single symmetric set, however tight, is therefore not evidence of a special arrangement; it
is the expected yield of choosing a few genes on a fly arm. The only symmetric sets that a
random control cannot produce are the tandem-scale ones, such as bcd--Dfd--Scr --- the very
ones a minimum-distance gate removes. This points to the real structure in the data: the
distance between neighbour genes, not its symmetry.

\subsection{The short-range population is the clustering signal}
\label{sec:control}

\begin{figure}[!htbp]
\centering
\includegraphics[width=\textwidth]{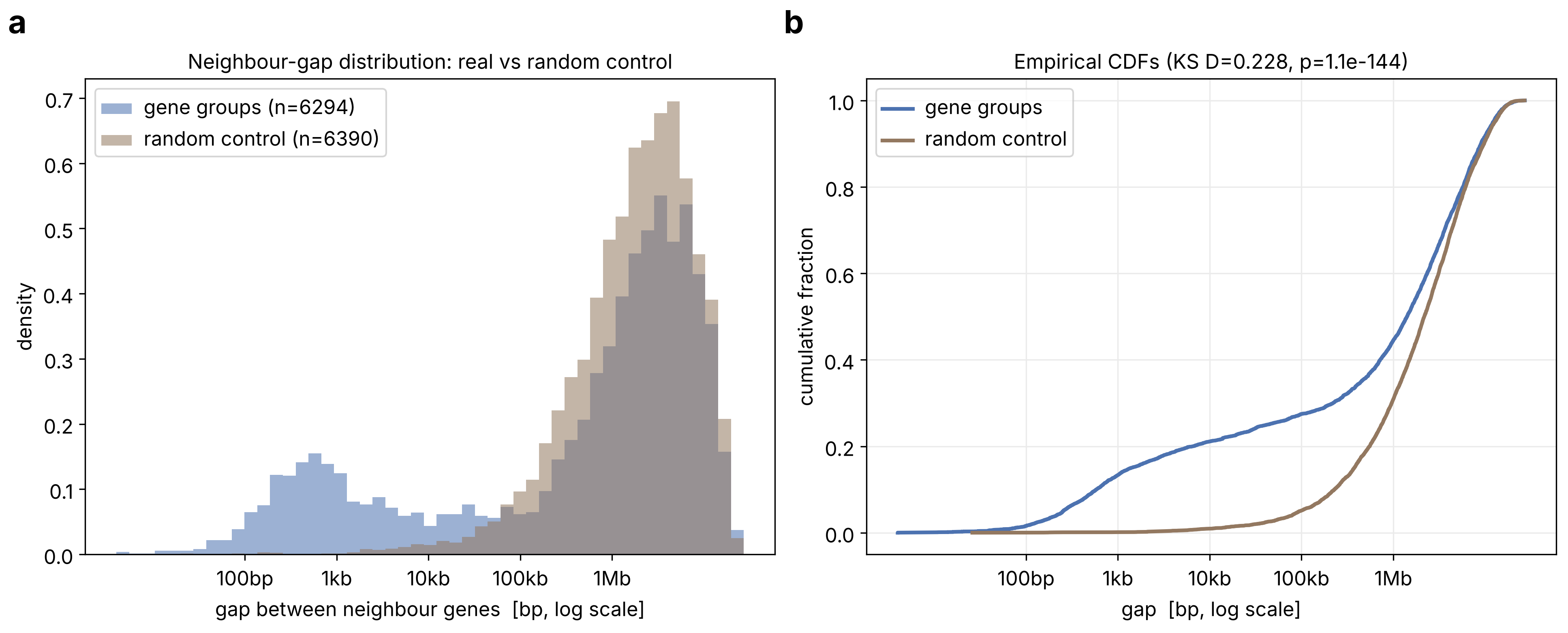}
\caption{Real fly gene groups against arm-matched random controls. (a)~Neighbour-distance
distributions: the long-range mode is shared; the short-range population (about 100~bp to
10~kb) is found only in the real groups. (b)~Empirical cumulative distributions (KS
$D = 0.228$, $p = 1.1\times10^{-144}$).}
\label{fig:control}
\end{figure}

For neighbour distances, the control shows how much of Fig.~\ref{fig:fly} is due to gene
families, and how much to the simple fact of choosing a few genes on a 25~Mb arm
(Fig.~\ref{fig:control}). Its distance distribution has one mode, at 1--5~Mb, that coincides
with the long-range mode of the real groups. The short-range population is absent: only 3 of
6,393 control pairs overlap (against 99 real pairs), and the control has essentially no
distances below 10~kb. A single gamma distribution fits the control almost exactly
($D = 0.015$), with shape 0.74--0.81 on the large arms. A shape near~1 is close to
exponential, the spacing law of random points. When a two-component mixture is forced on the
control, its `short' component sits at about 450~kb, the left shoulder of the single mode,
three orders of magnitude above the 1.8~kb short-range mode of the real groups. Real and
control distributions differ strongly (KS $D = 0.228$, $p \approx 10^{-144}$), and the
difference lies entirely in the short-range population, which holds about a quarter of the
real pairs.

The control also reproduces two features of the per-arm fits below: the near-identity of the
five large arms (\texttt{2L}--\texttt{3L}, \texttt{2L}--\texttt{3R} and
\texttt{3L}--\texttt{3R} are statistically indistinguishable in the control as in the real
data) and the outlier status of \texttt{4} (KS $D = 0.75$--0.80 against every arm). These
features follow from arm length and gene density, not from family biology. Two findings
survive the control: the short-range population, and the overlapping pairs.

\section{Two populations in eight genomes}
\label{sec:species}

\subsection{Gene positions on the ideograms of seven further genomes}
\label{sec:ideogram}

\begin{figure}[!htbp]
\centering
\includegraphics{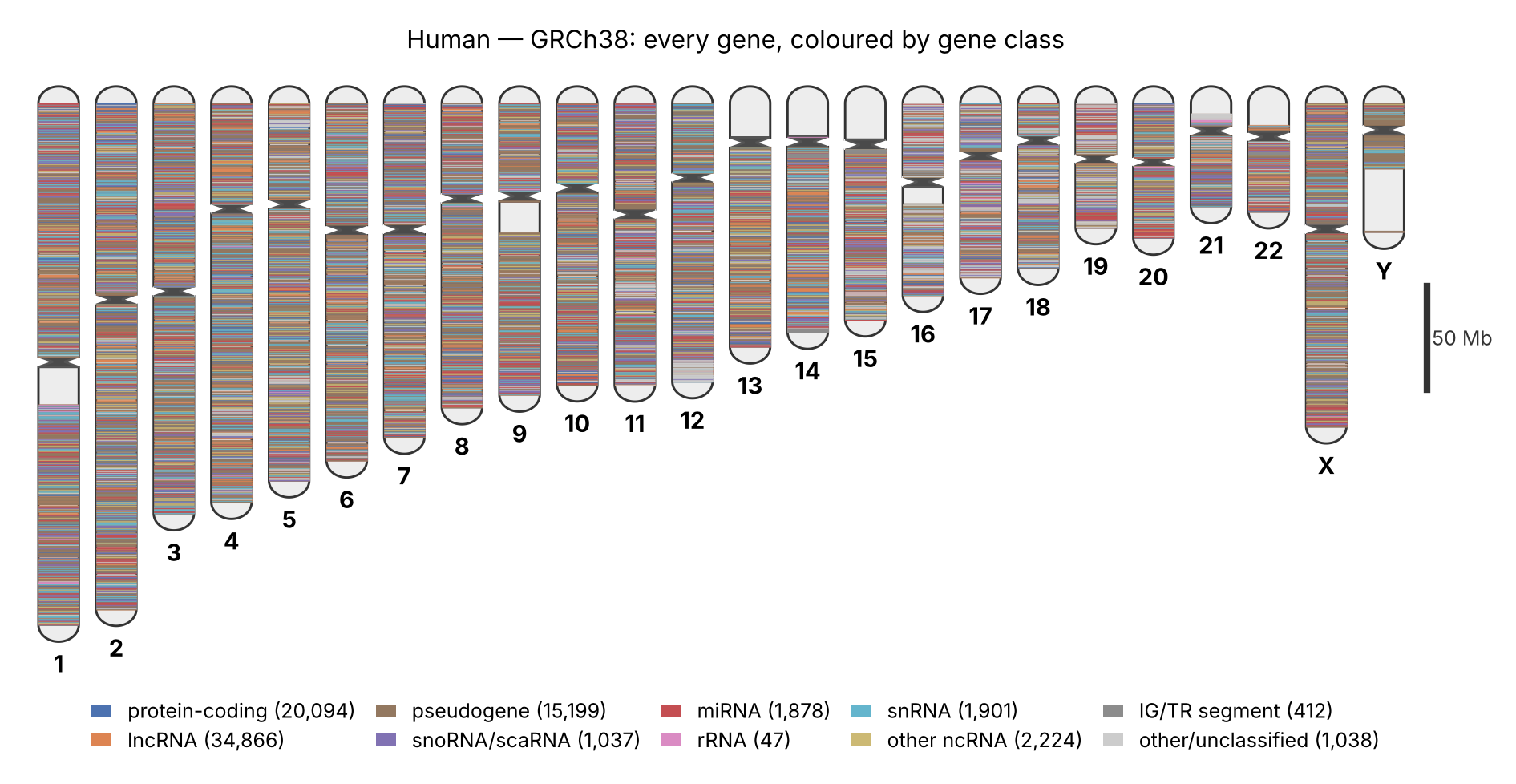}\\[6pt]
\includegraphics{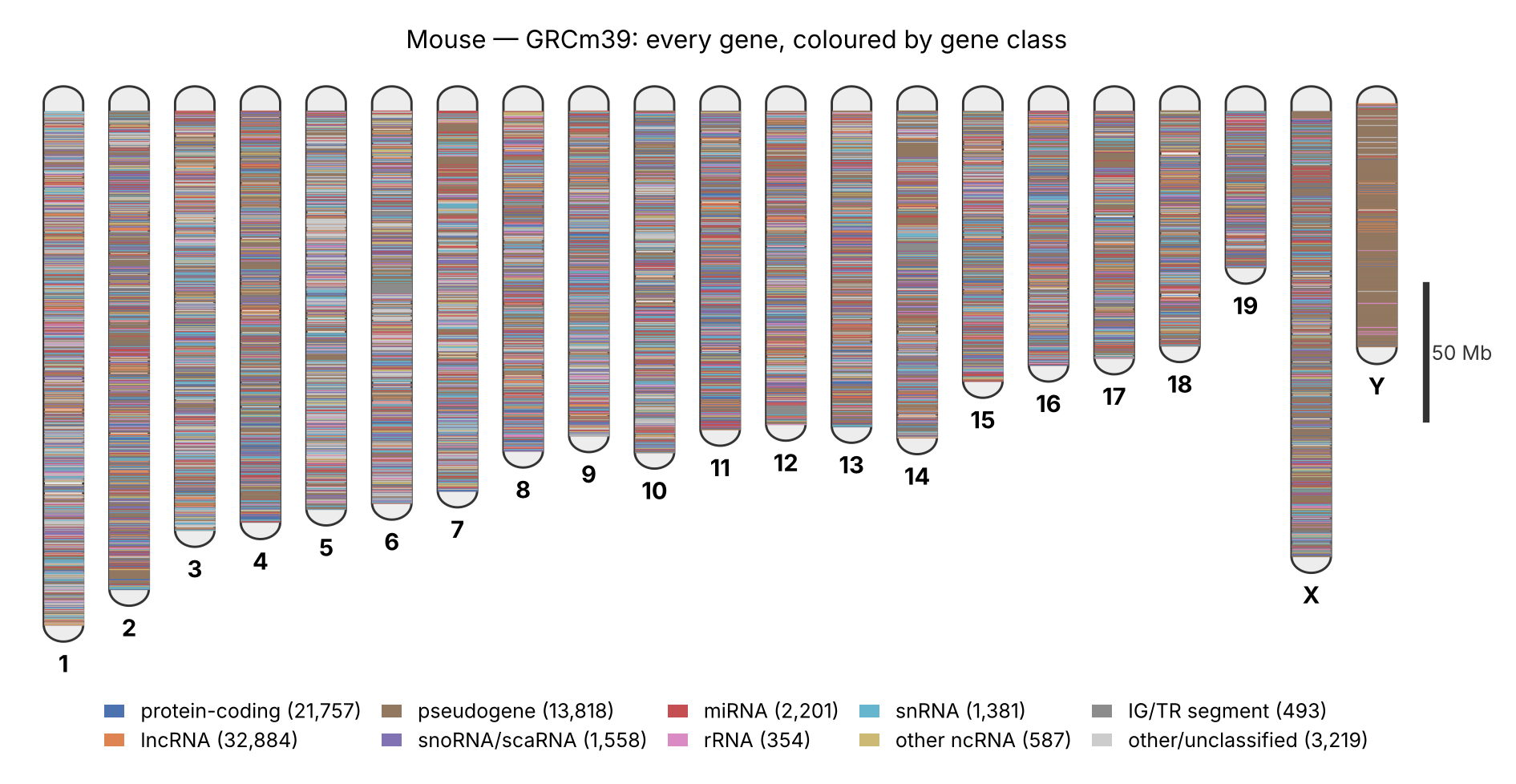}
\caption{Every located gene drawn at its position, coloured by gene class (classification as
in \cite{wu2026}): human (top; GRCh38, p and q arms joined at the centromere) and mouse
(bottom; GRCm39). Only fly has a band map in this study, so these chromosomes are drawn as plain bars
scaled to assembled length; chromosome widths and the tallest chromosome are drawn at least as
large as the fly ideograms of Fig.~\ref{fig:topsym}. Gene classes as in Fig.~\ref{fig:ideogram}b;
IG/TR, immunoglobulin and T-cell receptor gene segments. Numbers in the legend are gene counts.}
\label{fig:genomesA}
\end{figure}

\begin{figure}[!htbp]
\centering
\includegraphics{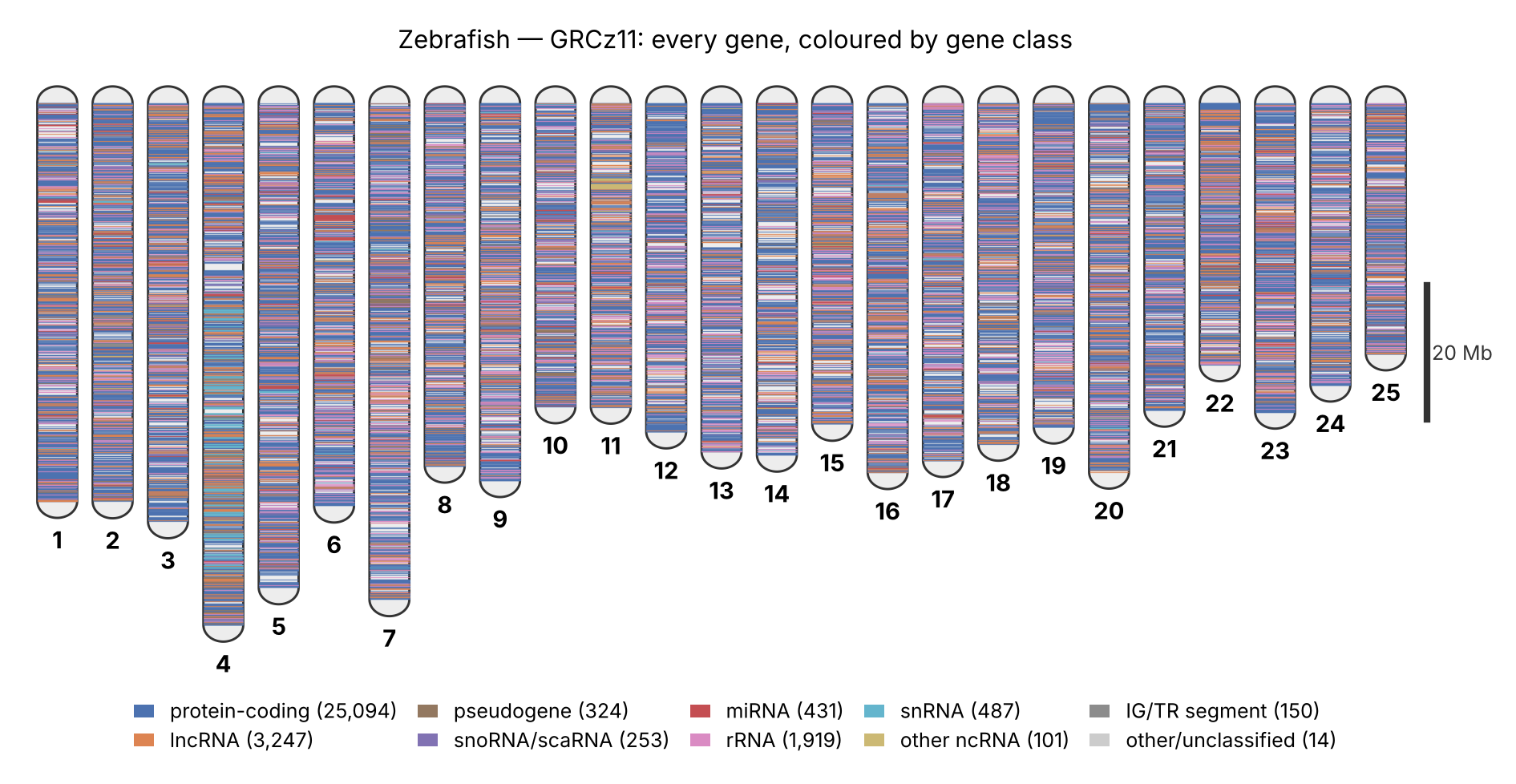}\\[6pt]
\includegraphics{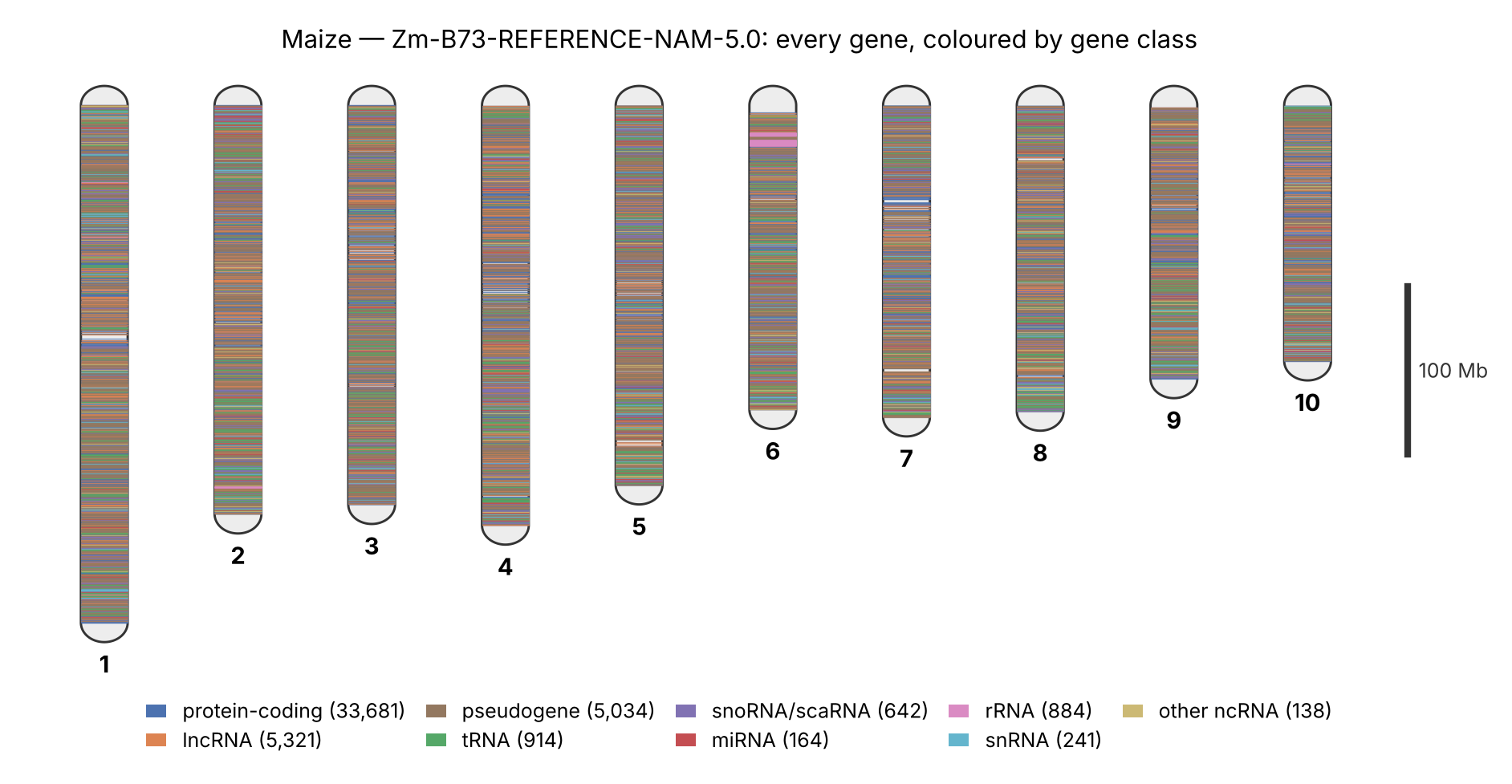}
\caption{As Fig.~\ref{fig:genomesA}, for zebrafish (top; GRCz11) and maize (bottom;
Zm-B73-REFERENCE-NAM-5.0).}
\label{fig:genomesB}
\end{figure}

\begin{figure}[!htbp]
\centering
\includegraphics{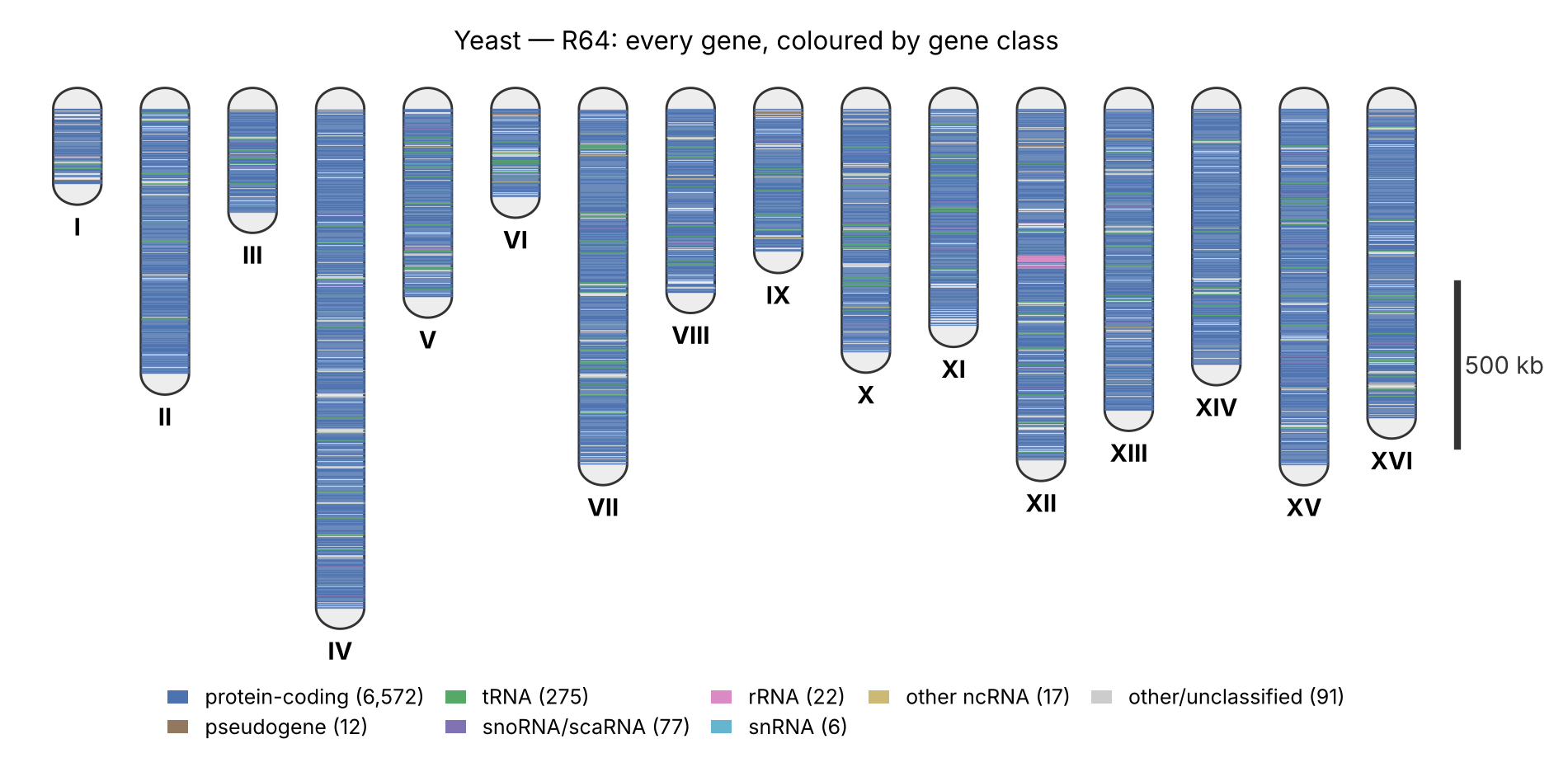}\\[6pt]
\includegraphics{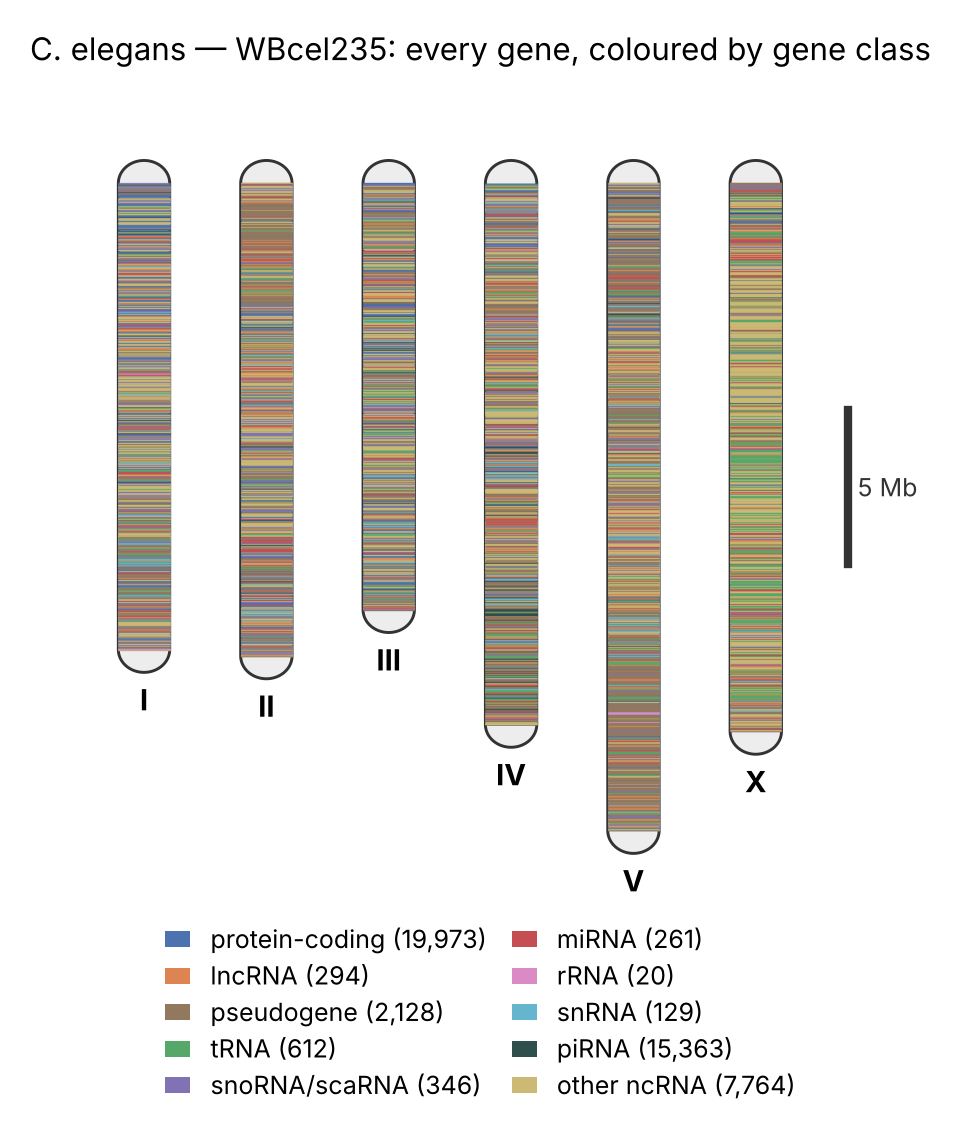}\hfill
\includegraphics{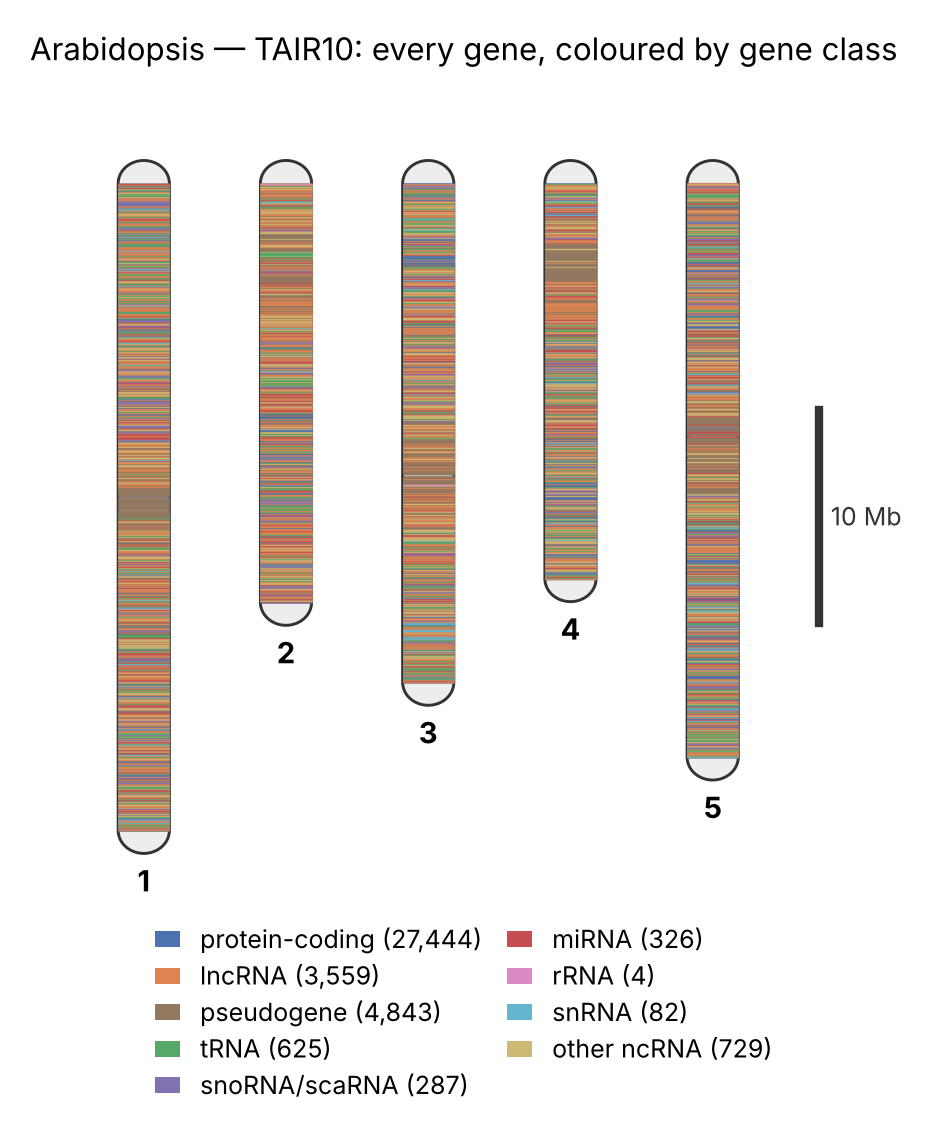}
\caption{As Fig.~\ref{fig:genomesA}, for yeast (top; \textit{S.~cerevisiae} R64-1-1),
\Ce{} (bottom left; WBcel235, holocentric chromosomes drawn without a centromere) and \At{}
(bottom right; TAIR10). In \Ce, 96\% of the 15,363 piRNA loci lie in two clusters on \texttt{IV}.}
\label{fig:genomesC}
\end{figure}

The whole-genome view of Fig.~\ref{fig:ideogram}b was drawn for human and the six other
model species (Figs.~\ref{fig:genomesA}--\ref{fig:genomesC}). It makes the differences in chromosome architecture that
matter later in the paper visible at a glance: the 48 human arms span more than a tenfold
range of length, from the gene-poor acrocentric short arms to \texttt{1p} and \texttt{2q};
mouse chromosomes are all acrocentric; the five \At{} chromosomes, the six \Ce{} chromosomes
and the ten maize chromosomes each vary in length by about a factor of two; and the sixteen
yeast chromosomes are all shorter than 1.6~Mb. In \Ce{}, 96\% of the 15,363 piRNA loci lie
in two clusters on \texttt{IV}, visible as two dense bands.

\subsection{The two populations recur in every genome}
\label{sec:speciesfits}

\begin{figure}[!htbp]
\centering
\includegraphics[width=0.49\textwidth]{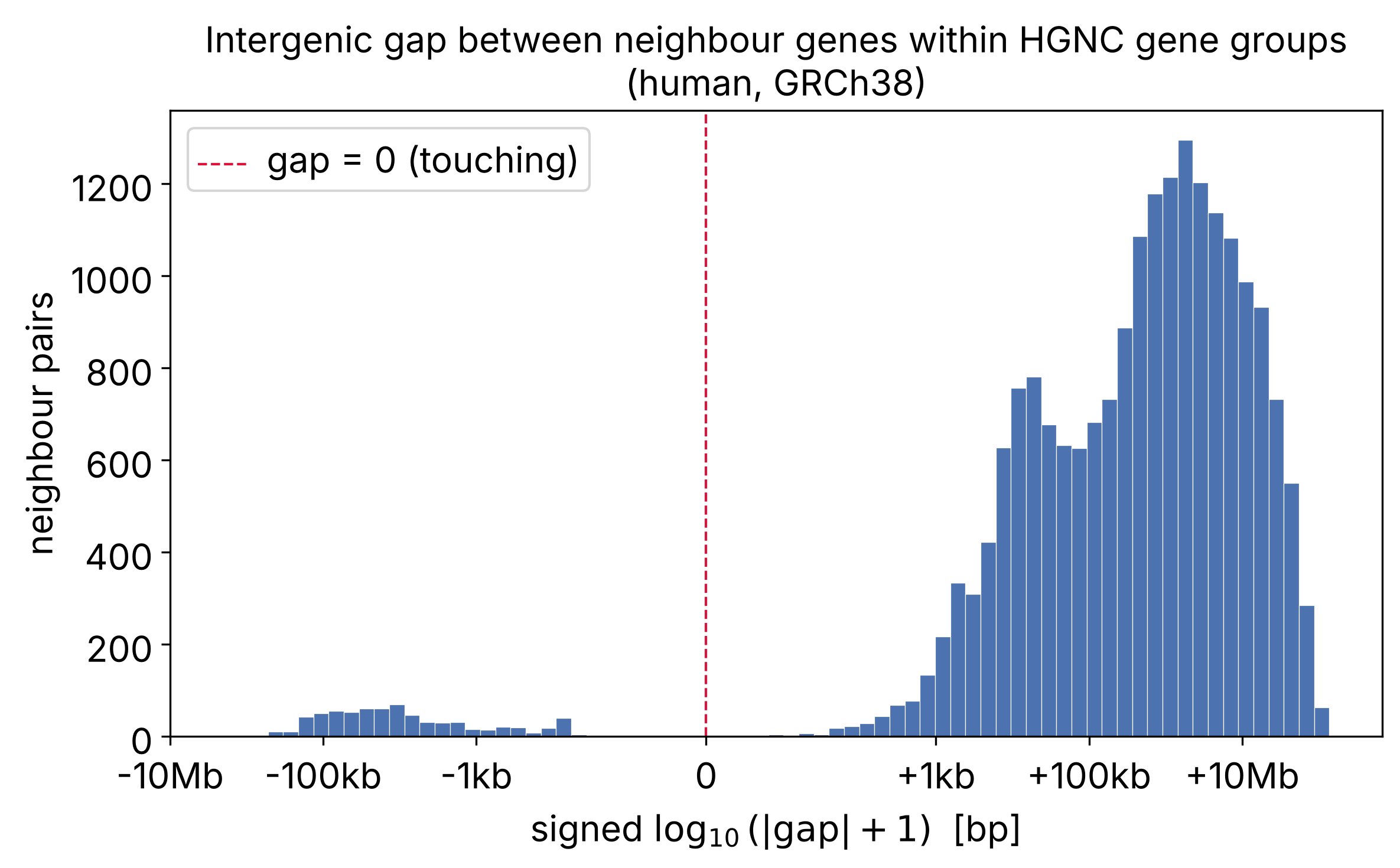}\hfill
\includegraphics[width=0.49\textwidth]{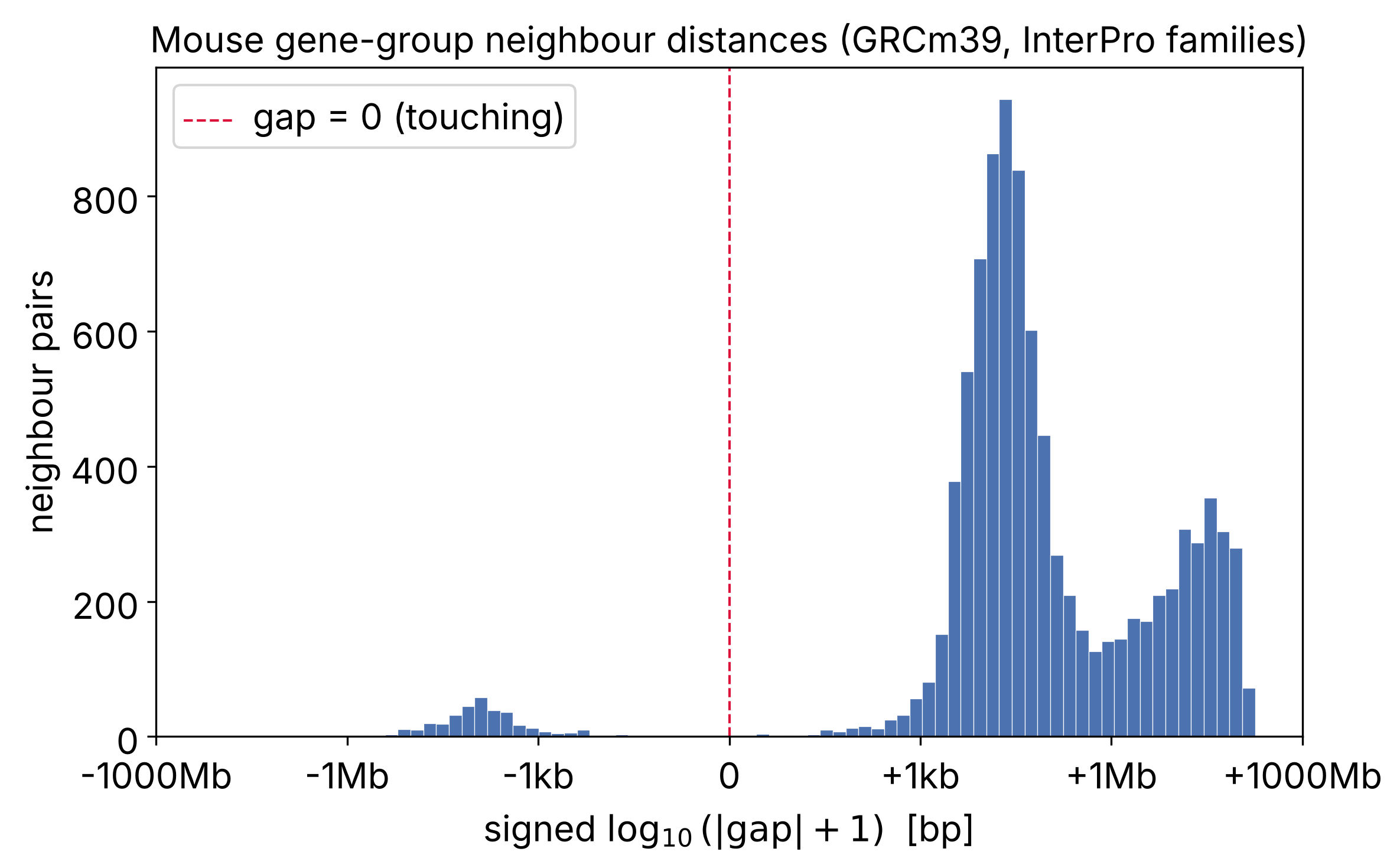}\\[4pt]
\includegraphics[width=0.49\textwidth]{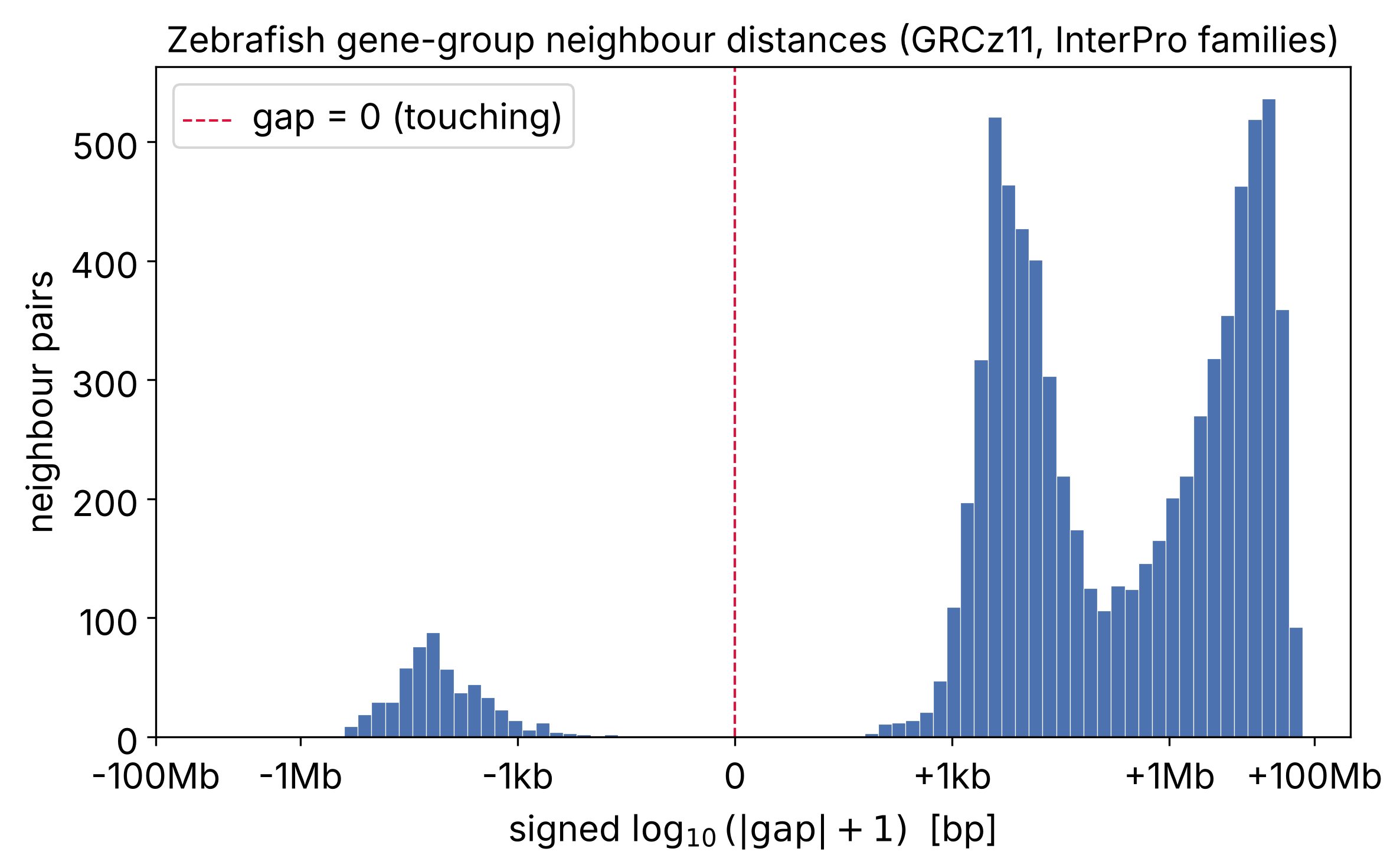}\hfill
\includegraphics[width=0.49\textwidth]{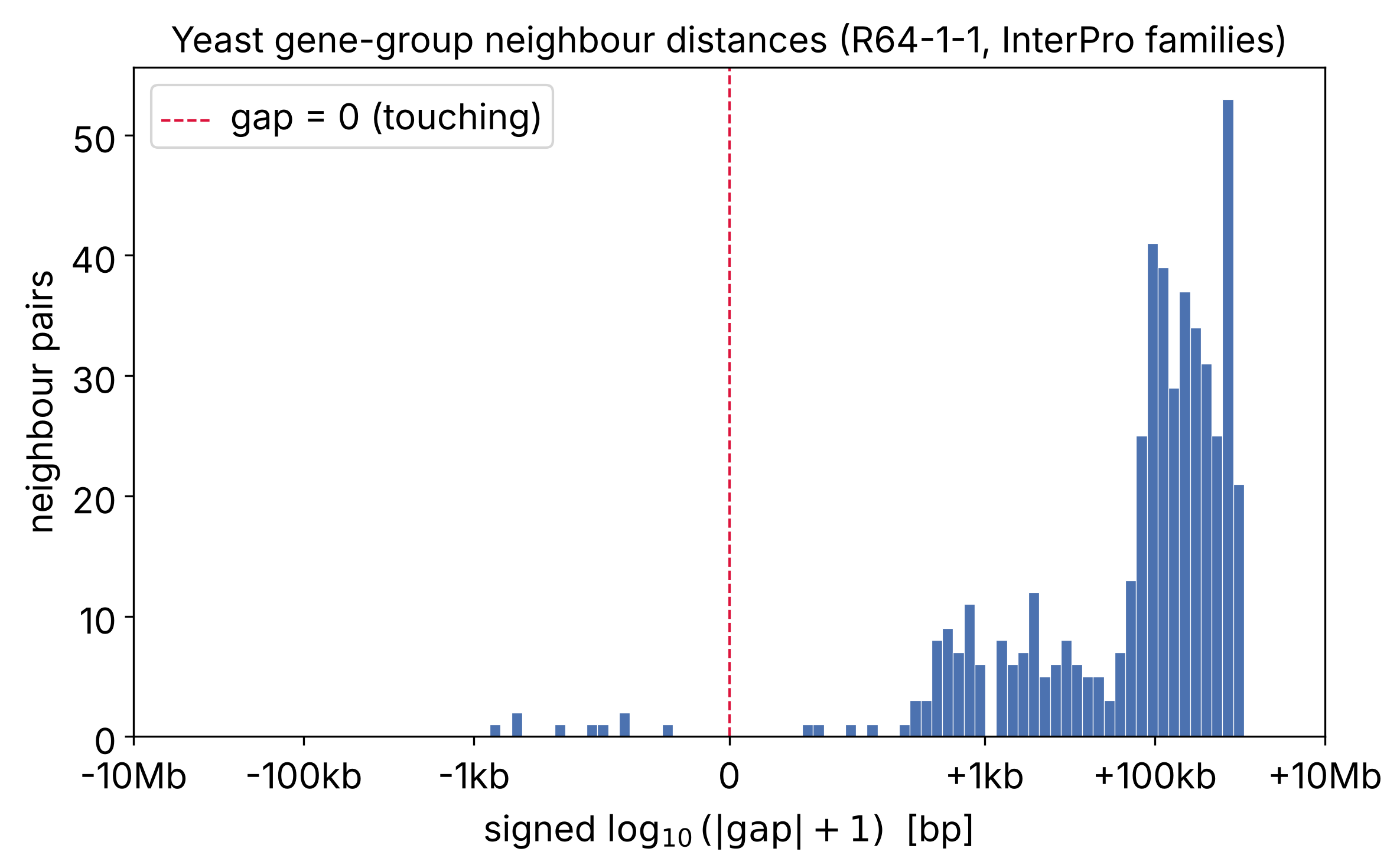}\\[4pt]
\includegraphics[width=0.49\textwidth]{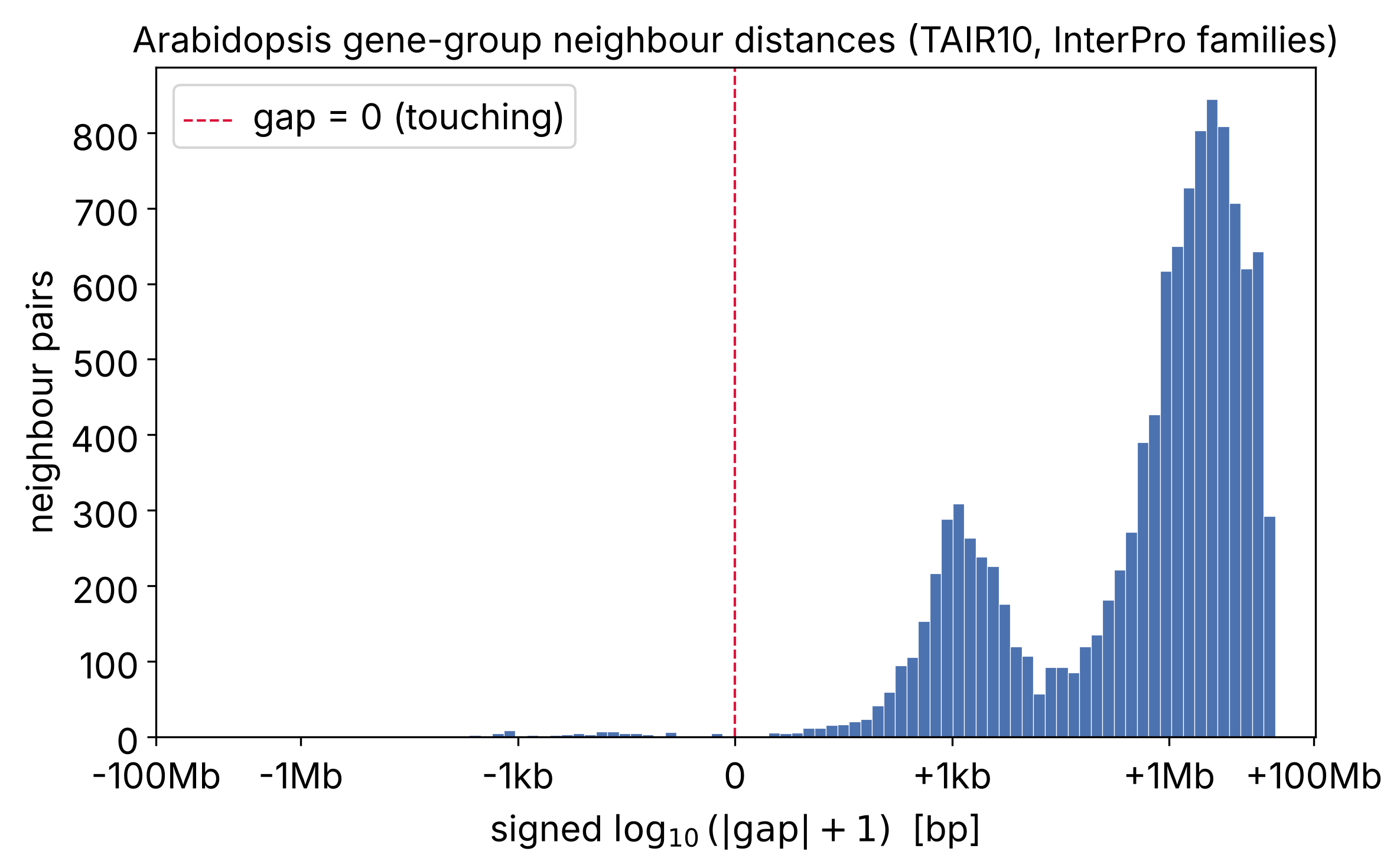}\hfill
\includegraphics[width=0.49\textwidth]{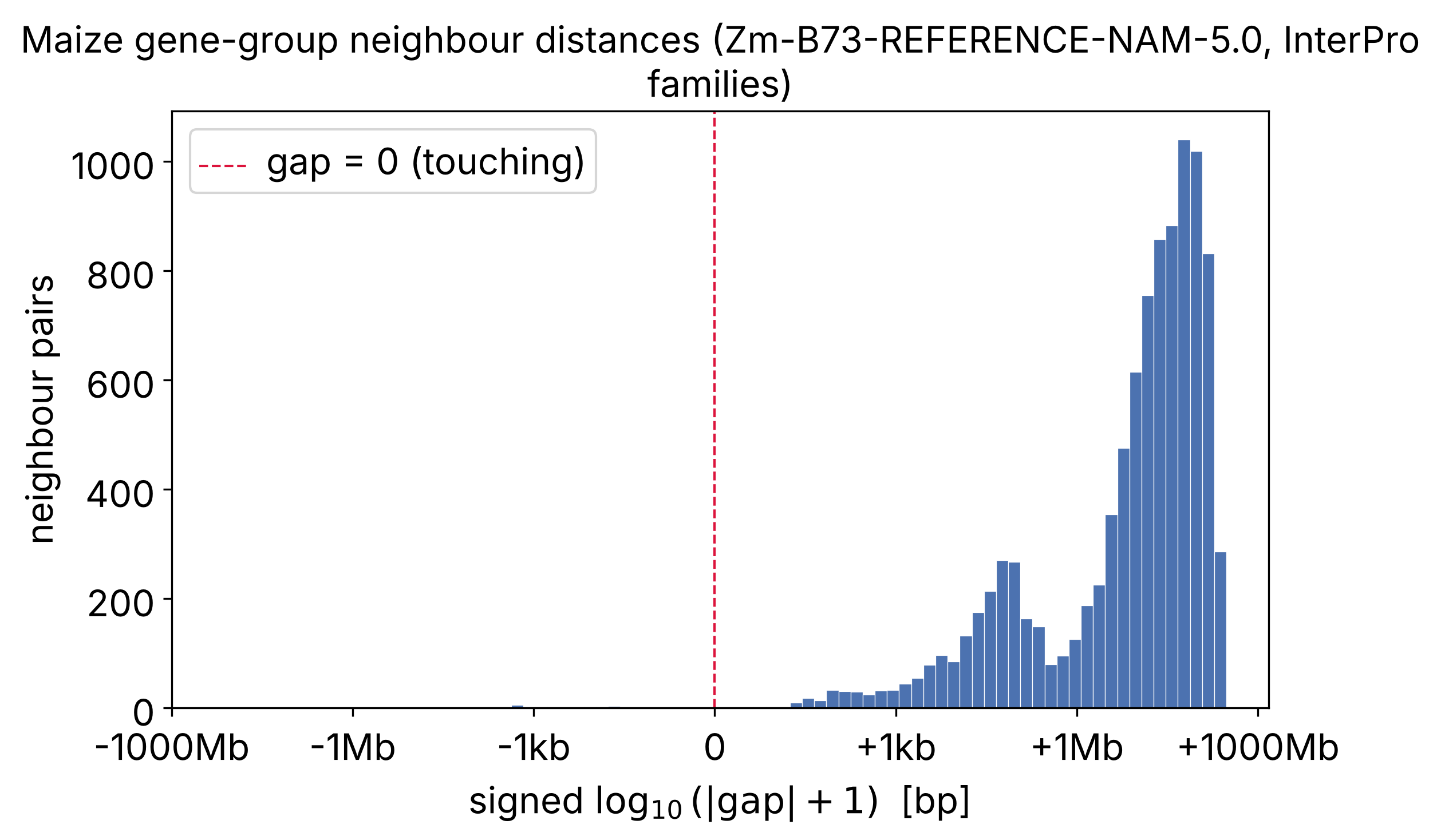}\\[4pt]
\includegraphics[width=0.49\textwidth]{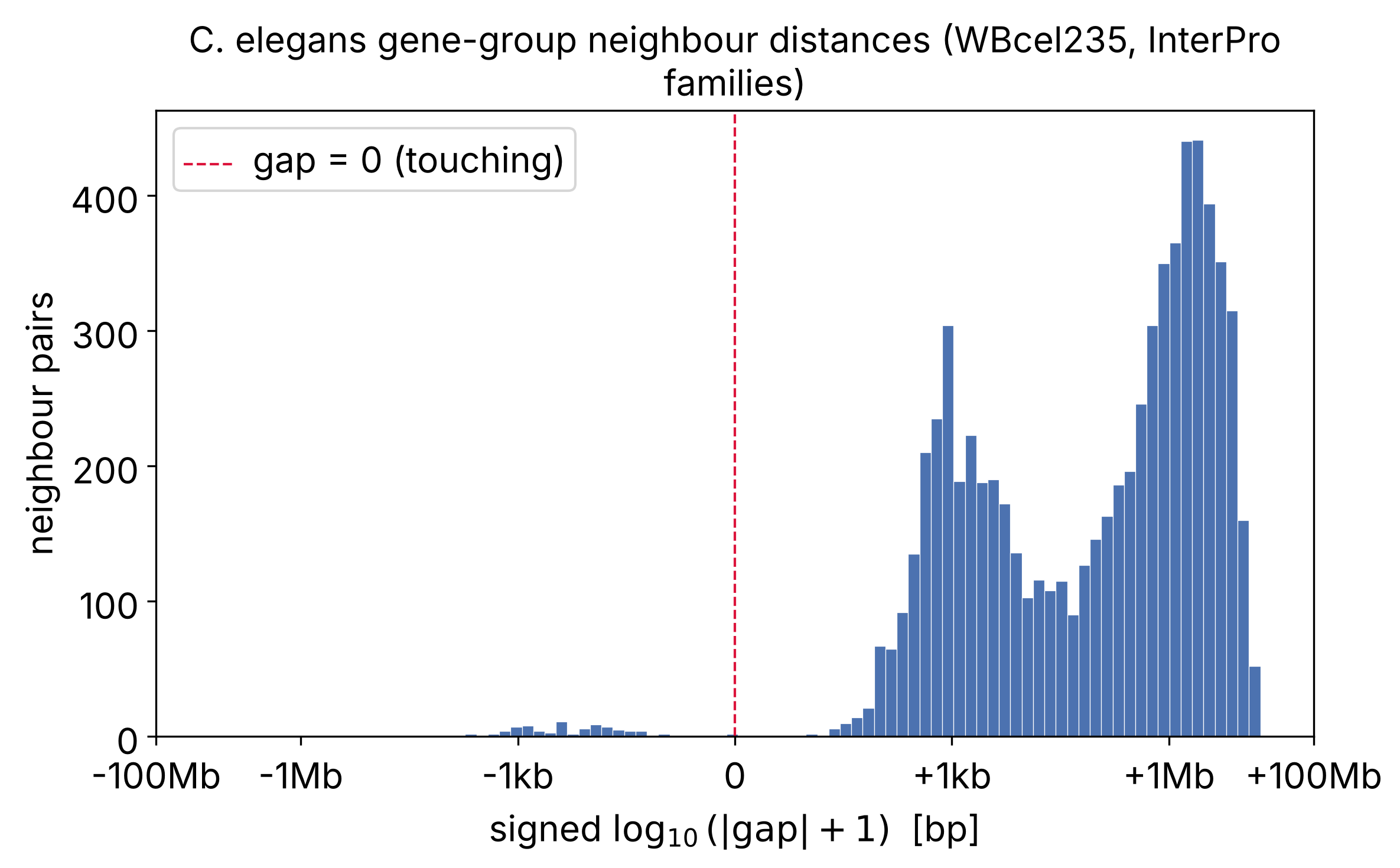}
\caption{Neighbour-distance distributions in seven further genomes, on the signed-log axis
of Fig.~\ref{fig:fly}a, with the axis range scaled to each genome. Left to right, top to
bottom: human (HGNC groups, p/q arms), mouse, zebrafish, yeast, \At, maize and \Ce{}
(InterPro \emph{Family} groups, whole chromosomes). Every genome shows a short-range and a
long-range population.}
\label{fig:species}
\end{figure}

The analysis was repeated on human, mouse, zebrafish, yeast, \At, maize and \Ce{}
(Fig.~\ref{fig:species}; Table~\ref{tab:summary}). The two-population shape is found in all
of them. The exponential null is rejected in every genome, most strongly in mouse
($D = 0.65$) and least strongly in yeast ($D = 0.18$). Among single distributions the
Weibull is best in human, the lognormal in mouse and the gamma in yeast, fly, \At, maize
and \Ce; in zebrafish gamma wins by AIC and lognormal by KS. As in fly, no single distribution
fits well, and in every genome the step from one to two mixture components gives by far the
largest BIC gain.

\begin{table}[!htbp]
\centering
\small
\caption{Neighbour distances in eight genomes: pooled two-component lognormal mixtures.
Overlap: fraction of pairs with $d \le 0$. $\Delta$BIC: improvement from $k = 1$ to $k = 2$.
The two weight columns are not comparable between rows (\S\ref{sec:weights}); the mode
columns are. Long-range modes are comparable only between genomes with the same unit.}
\label{tab:summary}
\begin{tabular}{@{}lrrrrrrrr@{}}
\toprule
Species & Pairs & Overlap & $\Delta$BIC & Exp.\ $D$ & Short mode & $\wshort$ & Long mode & Unit \\
\midrule
yeast       & 487    & 1.8\% & 266   & 0.18 & 1.77 kb & 0.26 & 230 kb  & chrom. \\
\Ce         & 7,115  & 1.2\% & 3,150 & 0.37 & 1.92 kb & 0.40 & 1.23 Mb & chrom. \\
\At         & 11,355 & 0.6\% & 7,050 & 0.25 & 1.90 kb & 0.26 & 2.49 Mb & chrom. \\
fly         & 6,393  & 1.5\% & 4,751 & 0.24 & 1.77 kb & 0.28 & 2.42 Mb & arm \\
zebrafish   & 7,909  & 6.9\% & 3,979 & 0.45 & 9.38 kb & 0.49 & 6.39 Mb & chrom. \\
maize       & 9,805  & 0.2\% & 6,027 & 0.21 & 69.2 kb & 0.28 & 33.3 Mb & chrom. \\
mouse       & 9,499  & 3.6\% & 4,149 & 0.65 & 20.7 kb & 0.71 & 14.0 Mb & chrom. \\
human       & 20,554 & 3.4\% & 2,157 & 0.38 & 22.9 kb & 0.37 & 2.96 Mb & arm \\
\bottomrule
\end{tabular}
\end{table}

The genomes differ in three ways that are visible in Table~\ref{tab:summary}.

\emph{Overlap.} The fraction of overlapping pairs ranges from 0.2\% (maize) and 0.6\%
(\At) to 3.4--3.6\% in the mammals and 6.9\% in zebrafish. Overlapping pairs are typically
genes nested within a neighbour's intron (median depth 7.8~kb in mouse, 14~kb in
zebrafish), not coincident loci. Zebrafish, with large tandem-expanded families such as
rhodopsin-like G-protein-coupled receptors (GPCRs; 731 genes), NOD-like (nucleotide-binding oligomerization domain-like) receptors (345) and
olfactory receptors (119), has the most.

\emph{Short-range mode.} It is 1.8--1.9~kb in yeast, fly, \At{} and \Ce{} (yeast and fly differ by less than
1~bp: 1,768 and 1,769~bp), 9.4~kb in
zebrafish, 21--23~kb in the two mammals and 69~kb in maize. That fly and \At, genomes of
almost the same size (144 and 135~Mb) separated by more than a billion years
\cite{timetree}, agree to within 8\% in the short-range mode (1.77 and 1.90~kb) and 3\% in
the long-range mode (2.42 and 2.49~Mb) was the first hint of what sets the scale (\S\ref{sec:h3}).

\emph{Weights.} Raw $\wshort$ ranges from 0.26 to 0.71. Section~\ref{sec:weights} shows
that most of this range comes from differences in group size.

\subsection{Robustness to the group definition}
\label{sec:robust}

Loosening the InterPro definition from \emph{Family} to \emph{Family}~+~\emph{Domain}
roughly doubles the number of groups and multiplies the pairs by 2.5--4, but leaves the
two-population shape intact in all InterPro genomes. The short-range mode moves by 7\% in
\Ce, by about 20\% in mouse, by 40\% in zebrafish and \At, and by a factor of two in maize
(69 to 140~kb); the maize mode is therefore best quoted as of order $10^5$~bp. In \Ce,
WormBase gene classes give a short-range mode of 1.98~kb and a long-range mode of 1.19~Mb,
within 3--4\% of the InterPro values. In human, removing the 29 biotype collections from the
HGNC groups removes about 45\% of the pairs and moves both modes (the short-range peak
sharpens near 10~kb and the long-range one shifts towards 10~Mb), but the distribution stays
bimodal.

\section{Two types of genome: consistent and variable per-chromosome mixtures}
\label{sec:types}

The pooled fits hide how the two populations behave chromosome by chromosome. Fitting the
mixture separately on every unit divides the eight genomes into two types
(Table~\ref{tab:types}; Figs.~\ref{fig:typeI} and \ref{fig:typeII}). In \emph{Type~I} genomes every chromosome or arm gives nearly the
same mixture: the short-range and long-range peaks sit in the same places and carry the same
weights, so the per-unit curves lie on top of one another. In \emph{Type~II} genomes the
per-unit mixtures fan out, in the position of the long-range peak, in the weight of the
short-range one, or in both.

\begin{table}[!htbp]
\centering
\small
\caption{Consistency of the per-unit two-component mixtures. Units: bimodal units fitted
(excluded as not bimodal or too small: fly \texttt{4}, human \texttt{5p}, mouse \texttt{Y},
yeast \texttt{XIII}). Ranges are over units; SD is the standard deviation over units of
$\log_{10}$ of the component median (dex) or of $\wshort$. Yeast: InterPro \emph{Family}
groups; with \emph{Family}~+~\emph{Domain} (12 chromosomes) the three SDs are 0.38, 0.10 and
0.073.}
\label{tab:types}
\setlength{\tabcolsep}{3.5pt}
\begin{tabular}{@{}llrrrrrrrr@{}}
\toprule
 & & & \multicolumn{2}{c}{Short-range median} & \multicolumn{2}{c}{Long-range median} & \multicolumn{2}{c}{$\wshort$} & Unit span \\
\cmidrule(lr){4-5}\cmidrule(lr){6-7}\cmidrule(lr){8-9}
Type & Species & Units & Range & SD & Range & SD & Range & SD & (Mb) \\
\midrule
I  & fly       & 5  & 1.3--3.7 kb  & 0.16 & 2.25--2.67 Mb & 0.03 & 0.22--0.30 & 0.027 & 23.5--32.1 \\
I  & \At       & 5  & 1.6--2.3 kb  & 0.06 & 1.81--3.29 Mb & 0.10 & 0.23--0.31 & 0.026 & 18.6--30.4 \\
I  & maize     & 10 & 47--186 kb   & 0.21 & 27.1--40.4 Mb & 0.06 & 0.23--0.39 & 0.055 & 152--308 \\
I  & \Ce       & 6  & 1.2--3.0 kb  & 0.13 & 0.72--1.91 Mb & 0.14 & 0.23--0.45 & 0.081 & 13.8--20.9 \\
\midrule
II & human     & 41 & 4.1--178 kb  & 0.41 & 0.57--9.46 Mb & 0.24 & 0.21--0.62 & 0.092 & 14.4--147 \\
II & mouse     & 20 & 11--124 kb   & 0.22 & 5.15--35.1 Mb & 0.20 & 0.48--0.83 & 0.097 & 57.9--192 \\
II & zebrafish & 25 & 5.1--23.9 kb & 0.16 & 1.14--12.3 Mb & 0.22 & 0.29--0.71 & 0.123 & 37.4--78.1 \\
II & yeast     & 8  & 1.0--5.7 kb  & 0.29 & 0.18--0.35 Mb & 0.08 & 0.13--0.41 & 0.107 & 0.75--1.53 \\
\bottomrule
\end{tabular}
\end{table}

\subsection{Type I: fly, maize, \At{} and \Ce}

\begin{figure}[!htbp]
\centering
\includegraphics[width=0.98\textwidth]{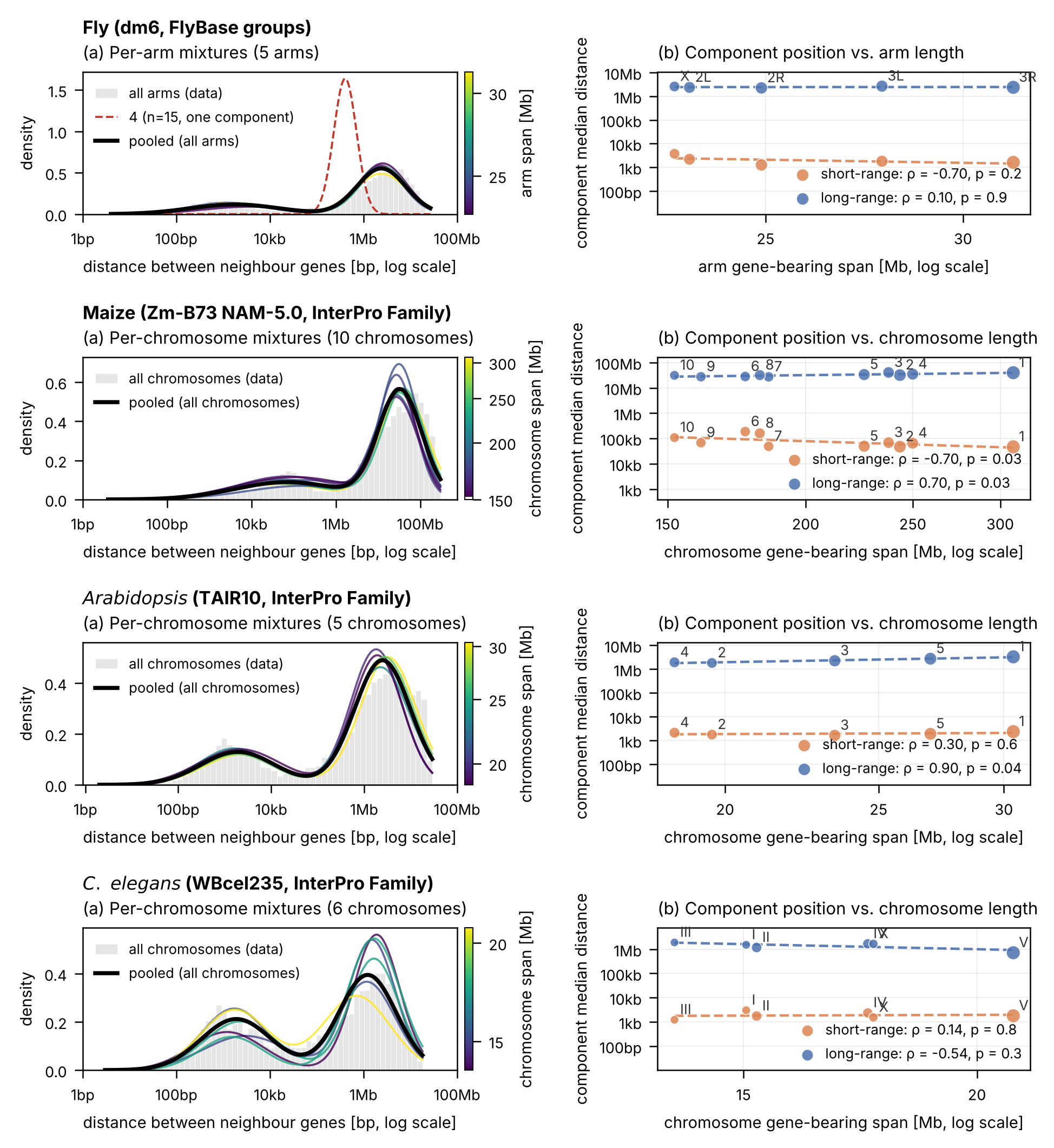}
\caption{Type~I genomes: consistent per-unit mixtures. Rows, top to bottom: fly (dm6, FlyBase
groups, arms), maize, \At{} and \Ce{} (InterPro \emph{Family} groups, whole chromosomes).
Units fitted, all bimodal: fly 5 large arms, maize 10, \At{} 5 and \Ce{} 6 chromosomes. In
each row, (a)~two-component lognormal mixtures fitted separately on each unit with at least 50
positive distances, coloured by the unit's gene-bearing span, over the pooled data (grey) and
the pooled fit (black); (b)~the short-range (orange) and long-range (blue) component medians
of each unit against its span, with log--log regression lines (dashed), Spearman $\rho$ in the
legend and marker size proportional to the number of pairs. Fly: the five large arms
coincide, and \texttt{4} (15 pairs) is fitted with one component (red dashed). Maize:
long-range $\rho = 0.70$, short-range $\rho = -0.70$. \At: long-range $\rho = 0.90$,
short-range $\rho = 0.30$. \Ce: the peaks are consistent, but the short-range weight varies
(0.23--0.45) and chromosome~\texttt{V} has the lowest long-range median.}
\label{fig:typeI}
\end{figure}

Fly is the clearest Type~I genome (Fig.~\ref{fig:typeI}, top row). The five large arms give
short-range medians of 1.3--3.7~kb with weights 0.22--0.30, and long-range medians of
2.25--2.67~Mb (SD 0.03~dex), and their fitted curves almost coincide. Pairwise KS distances
between the large arms are small ($D = 0.04$--0.10), and several pairs, such as
\texttt{2L}--\texttt{3L} and \texttt{3L}--\texttt{3R}, are statistically indistinguishable.
Chromosome~\texttt{4} (15 pairs) is the only outlier, fitted with a single component, because
a 1.3~Mb chromosome cannot hold megabase distances.

Maize, a genome fifteen times larger, behaves the same way (Fig.~\ref{fig:typeI}, second
row). All ten
chromosomes are bimodal, the long-range medians lie between 27 and 40~Mb (SD 0.06~dex), and
the short-range weights between 0.23 and 0.39. The short-range medians spread more
(47--186~kb), driven by chromosomes \texttt{6} and \texttt{8}, but the curves still share
one shape. The long-range median rises gently with chromosome span ($\rho = 0.70$,
$p = 0.025$; slope 0.49), as random placement predicts.

\At{} is as consistent as fly (Fig.~\ref{fig:typeI}, third row): all five chromosomes are
bimodal, with short-range medians of 1.6--2.3~kb (SD 0.06~dex, the tightest of all eight
genomes), long-range medians of 1.8--3.3~Mb that track chromosome span ($\rho = 0.90$,
$p = 0.04$), and weights of 0.23--0.31. \Ce{} is Type~I at the edge of the definition
(Fig.~\ref{fig:typeI}, bottom row): the short-range medians (1.2--3.0~kb) and long-range
medians (0.7--1.9~Mb) are consistent, but the short-range weight varies from 0.23 on
\texttt{X} to 0.45 on \texttt{II} and \texttt{V}, and chromosome~\texttt{V}, which holds 38\%
of all pairs, has the lowest long-range median (0.72~Mb). Both deviations follow from the
tandem-arrayed chemoreceptor families that crowd the arms of the autosomes, above all
\texttt{V} \cite{robertson} (\S\ref{sec:h3}).

What the four Type~I genomes share is that their units are of similar size. The five large
fly arms span 23.5--32.1~Mb, the \At{} chromosomes 18.6--30.4~Mb, the \Ce{} chromosomes
13.8--20.9~Mb and the maize chromosomes 152--308~Mb: no more than a twofold range in any of
them. Since the long-range population is random placement (\S\ref{sec:control}), units of
similar length give long-range peaks in similar places.

\subsection{Type II: human, mouse, zebrafish and yeast}

\begin{figure}[!htbp]
\centering
\includegraphics[width=\textwidth]{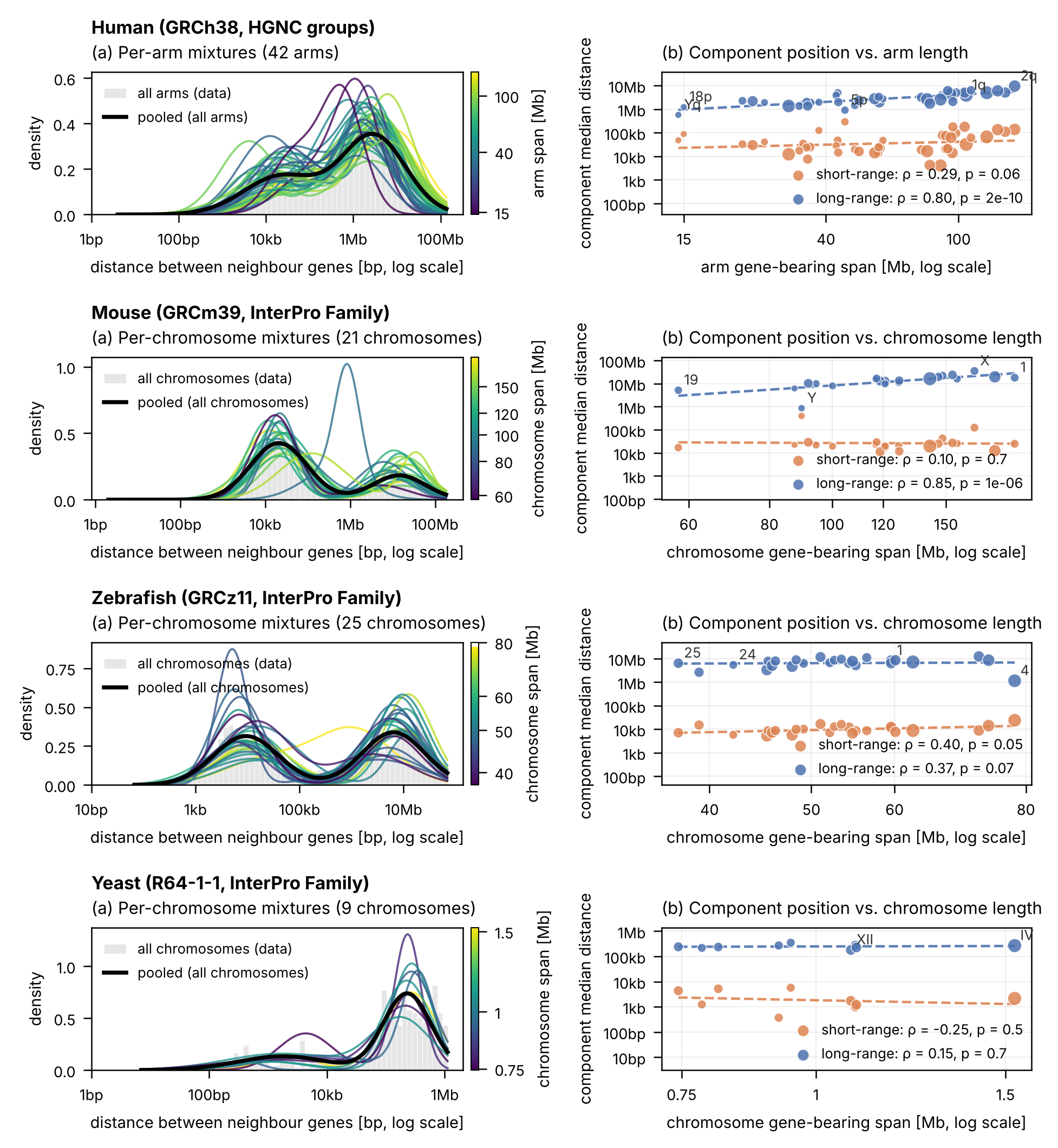}
\caption{Type~II genomes: variable per-unit mixtures. Rows, top to bottom: human (GRCh38, HGNC
groups, p/q arms), mouse, zebrafish and yeast (InterPro \emph{Family} groups, whole
chromosomes). In each row, (a)~and~(b) as in Fig.~\ref{fig:typeI}. Human: 42 arms (41 bimodal; \texttt{5p} is not); the
long-range median scales with arm span ($\rho = 0.80$), the short-range median does not
($\rho = 0.29$). Mouse: 21 chromosomes (20 bimodal; \texttt{Y} is not); long-range $\rho = 0.85$, short-range $\rho = 0.10$.
Zebrafish: 25 chromosomes (all bimodal) whose spans vary only twofold, yet the short-range weights range from
0.29 to 0.71 and chromosome~\texttt{4} has a long-range median of 1.1~Mb against 2.6--12~Mb
elsewhere. Yeast: the nine chromosomes with at least 25 positive distances (eight bimodal;
\texttt{XIII} is not); the
\emph{Family}~+~\emph{Domain} groups are shown in Appendix Fig.~\ref{fig:yeastfd}.}
\label{fig:typeII}
\end{figure}

Human is the clearest Type~II genome (Fig.~\ref{fig:typeII}, top row). Of the 42 arms with at least
50 positive distances, 41 are bimodal (\texttt{5p} is the exception); the four acrocentric short arms
(\texttt{13p}, \texttt{14p}, \texttt{15p}, \texttt{22p}) carry no group members at all. The
per-arm curves fan out across two orders of magnitude: long-range medians run from 0.57~Mb
(\texttt{Yq}) to 9.5~Mb (\texttt{2q}), short-range medians from 4~kb (\texttt{14q},
\texttt{15q}) to 178~kb (\texttt{3q}), and short-range weights from 0.21 (\texttt{16q}) to
0.62 (\texttt{19q}). Part of the spread is geometry: human arms differ more than tenfold in
length, and the long-range median climbs with arm span (Spearman $\rho = 0.80$,
$p = 2\times10^{-10}$; log--log slope 0.77, $R^2 = 0.62$), whereas the short-range median
does not ($\rho = 0.29$, $p = 0.06$; $R^2 = 0.03$; Fig.~\ref{fig:typeII}, top row,~b). Part is
family biology: the highest short-range weights are on \texttt{19p} and \texttt{19q}
(0.53, 0.62), home to the largest clusters of KRAB (Kr\"uppel-associated box) zinc-finger genes \cite{huntley}.

Mouse (Fig.~\ref{fig:typeII}, second row) shows the same scaling more cleanly: 20 of 21
chromosomes are bimodal (all but \texttt{Y}), the long-range median scales with chromosome
span ($\rho = 0.85$, $p = 10^{-6}$; 5.2--35~Mb), and the short-range median is flat
($R^2 = 0.00$; autosomal medians 11--42~kb across chromosomes of 58--192~Mb). The weights
vary more than in any Type~I genome (0.48--0.83), and \texttt{X} stands out with a
short-range median of 124~kb.

Zebrafish (Fig.~\ref{fig:typeII}, third row) shows that Type~II is not only a matter of unit length.
Its 25 chromosomes vary in span by only a factor of two (37--78~Mb), less than maize's, and
the long-range median does not scale with span ($\rho = 0.37$, $p = 0.07$). Yet the
short-range weight ranges from 0.29 to 0.71 --- the largest SD of all eight genomes --- and
chromosome~\texttt{4}, whose long arm carries large tandem gene clusters \cite{howe}, has a
long-range median of 1.1~Mb, two to ten times below the others. The variation here comes
from where the tandem-expanded families lie, not from chromosome geometry.

Yeast (Fig.~\ref{fig:typeII}, bottom row) is Type~II for a different reason. The long-range
medians are consistent (0.18--0.35~Mb), but each chromosome holds only 26--103 pairs, and
the short-range component is fitted on a handful of them: its median ranges from 1.0 to
5.7~kb and its weight from 0.13 to 0.41, and chromosome~\texttt{XIII} is not bimodal. With
the larger \emph{Family}~+~\emph{Domain} set (12 chromosomes; Appendix Fig.~\ref{fig:yeastfd})
the weights become more
consistent but the short-range medians spread further (0.5--14~kb), and the long-range
median now scales with chromosome length ($\rho = 0.65$, $p = 0.02$). With the
\emph{Family} set alone only nine chromosomes can be fitted and no trend is detectable.
Yeast's variability is thus partly a sampling limit of a small genome.

\subsection{What the two types share}
\label{sec:longrange}

Across both types, the long-range population behaves exactly as random placement predicts:
the distance between randomly chosen genes grows with the room available. The long-range
median rises significantly with unit span in human and mouse, whose units span a more than
threefold range, and also in maize, \At{} and yeast with \emph{Family}~+~\emph{Domain},
where the range is only about twofold and the trend is shallow (slope 0.49 in maize). It does
not in \Ce{} ($\rho = -0.54$, $p = 0.3$), whose six chromosomes span only a 1.5-fold range,
or in zebrafish ($\rho = 0.37$, $p = 0.07$), where chromosome~\texttt{4} is an outlier
(\S\ref{sec:types}). Within a twofold range the long-range peak moves little, which is why
such genomes can still be Type~I. In yeast, randomly placed members of a family with 2--3 members on a 0.9~Mb
chromosome would be about 200--300~kb apart, which matches the fitted 230~kb, and it explains
why the exponential null is only barely rejected there. Long-range modes can therefore be
compared only between genomes with the same unit. Mouse's 14~Mb and maize's 33~Mb reflect
whole-chromosome units of 58--308~Mb, not a property of families, and the agreement between
fly (2.4~Mb) and human (3.0~Mb) is a coincidence of arm lengths.

The short-range median, in contrast, does not follow unit span in any genome. It is nearly
constant across the units of every Type~I genome, and in Type~II genomes it varies without
regard to span. Its scale is a property of the genome as a whole, which the next section
explains.

\section{The short-range scale is set by gene spacing}
\label{sec:h3}

\begin{figure}[!htbp]
\centering
\includegraphics[width=\textwidth]{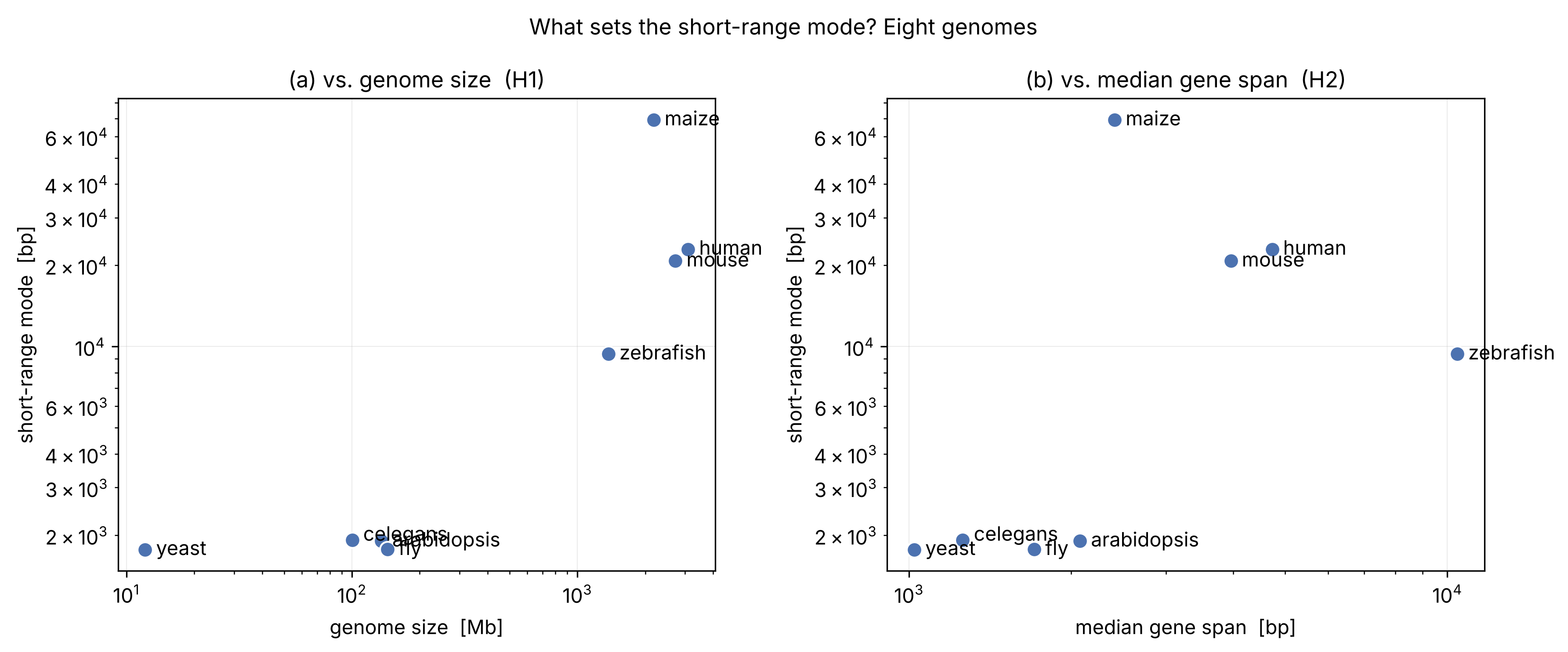}\\[4pt]
\includegraphics[width=0.55\textwidth]{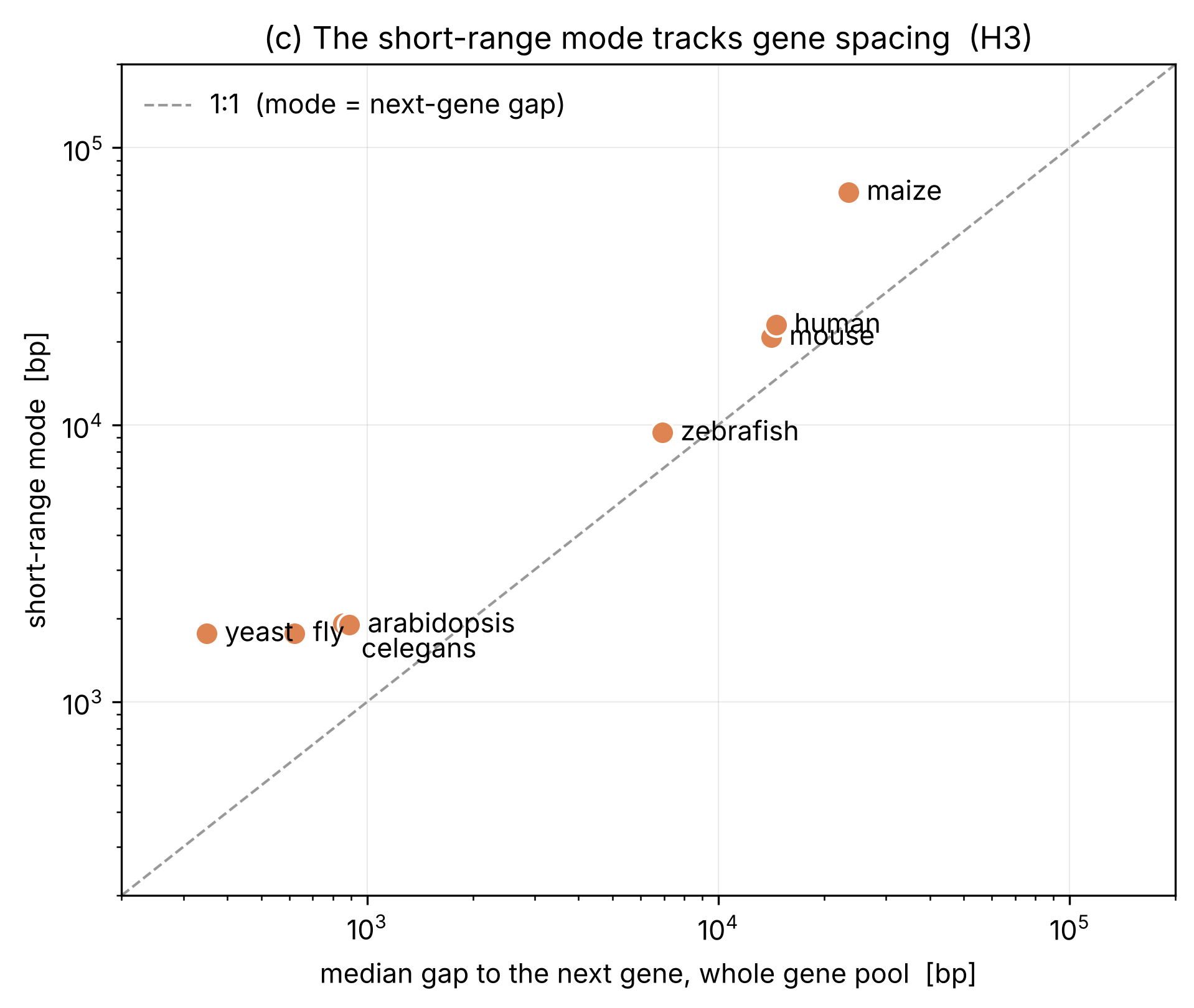}
\caption{What sets the short-range mode? Short-range mode of each genome against (a)~genome
size (H1), (b)~median gene span (H2), and (c)~median distance from one gene to the next in
the whole gene pool (H3), with the 1:1 line dashed. Gene spacing gives by far the closest
rank agreement (Spearman $\rho = 0.98$; only \Ce{} and \At{} swap places).}
\label{fig:h3}
\end{figure}

The short-range population is the genuine clustering signal, so its scale is the quantity
worth explaining. Three hypotheses were tested (Fig.~\ref{fig:h3}; Table~\ref{tab:h3}).

\emph{H1: genome size.} The short-range mode increases with genome size ($\rho = 0.83$,
$p = 0.010$; log--log $R^2 = 0.77$), but slowly (slope 0.63), and with clear exceptions:
maize, at 70\% of the size of the human genome, has a mode three times larger, and
zebrafish lies well below the trend.

\emph{H2: gene size.} If a tandem copy must clear the span of the gene it copies, the mode
should follow gene length. It does not ($\rho = 0.71$, $p = 0.047$; $R^2 = 0.33$). The
two exceptions come from opposite directions. Maize genes are as compact as \At{} genes
(median span 2.4 vs 2.1~kb), but its mode is 29 times its median gene; zebrafish has the
longest genes in the set (median 10.4~kb, more than twice human's), but a mode below both
mammals'.

\emph{H3: gene spacing.} What maize has that \At{} lacks, and what zebrafish lacks relative
to the mammals, is DNA between adjacent genes. The median next-gene distance explains the
mode across all eight genomes ($\rho = 0.98$, $p < 0.001$; $R^2 = 0.94$, slope 0.84;
Fig.~\ref{fig:h3}c). Expressed as a multiple of the next-gene distance, the mode varies only
from 1.35 to 5.07 (0.57~dex), compared with a 39-fold (1.59~dex) range in the raw modes;
gene spacing thus accounts for about two thirds of the variation between genomes. Every
genome lies above the 1:1 line: a short-range pair is typically one to a few genes apart, not
strictly adjacent.

\begin{table}[!htbp]
\centering
\small
\caption{Short-range mode, gene size and gene spacing. Next-gene distance: median distance
between consecutive genes in the whole gene pool (overlapping pairs excluded; \Ce{} piRNA
loci excluded). Overlapping: fraction of adjacent gene pairs that overlap.}
\label{tab:h3}
\setlength{\tabcolsep}{4pt}
\begin{tabular}{@{}lrrrrrr@{}}
\toprule
 & Genome & Gene span & Next-gene & & Short mode & Mode / \\
Species & (Mb) & (bp) & dist.\ (bp) & Overlapping & (bp) & dist. \\
\midrule
yeast       & 12.1  & 1,023  & 349    & 0.10 & 1,768  & 5.07 \\
\Ce         & 100.3 & 1,259  & 855    & 0.21 & 1,919  & 2.25 \\
\At         & 135.0 & 2,078  & 891    & 0.13 & 1,902  & 2.14 \\
fly         & 143.7 & 1,709  & 622    & 0.21 & 1,769  & 2.84 \\
zebrafish   & 1,373 & 10,433 & 6,926  & 0.15 & 9,383  & 1.35 \\
maize       & 2,178 & 2,410  & 23,468 & 0.10 & 69,196 & 2.95 \\
mouse       & 2,723 & 3,961  & 14,128 & 0.29 & 20,748 & 1.47 \\
human       & 3,100 & 4,727  & 14,603 & 0.34 & 22,938 & 1.57 \\
\midrule
\multicolumn{7}{@{}l}{Spearman $\rho$ with the short mode: genome size 0.83; gene span 0.71;
next-gene distance 0.98.} \\
\bottomrule
\end{tabular}
\end{table}

Two genomes were not used to formulate H3 and serve as out-of-sample checks. Zebrafish, with
a next-gene distance of 6.9~kb, is the only genome between the small-genome cluster
(0.35--0.9~kb) and the large-genome cluster (14--23~kb), and it falls on the relationship.
\Ce{} was predicted, before fitting, to have a mode of a few kilobases from its 855~bp
next-gene distance; the fitted mode is 1.92~kb (ratio 2.25). It falls inside an already
populated cluster, so it is a weaker test.

\begin{figure}[!htbp]
\centering
\includegraphics[width=0.75\textwidth]{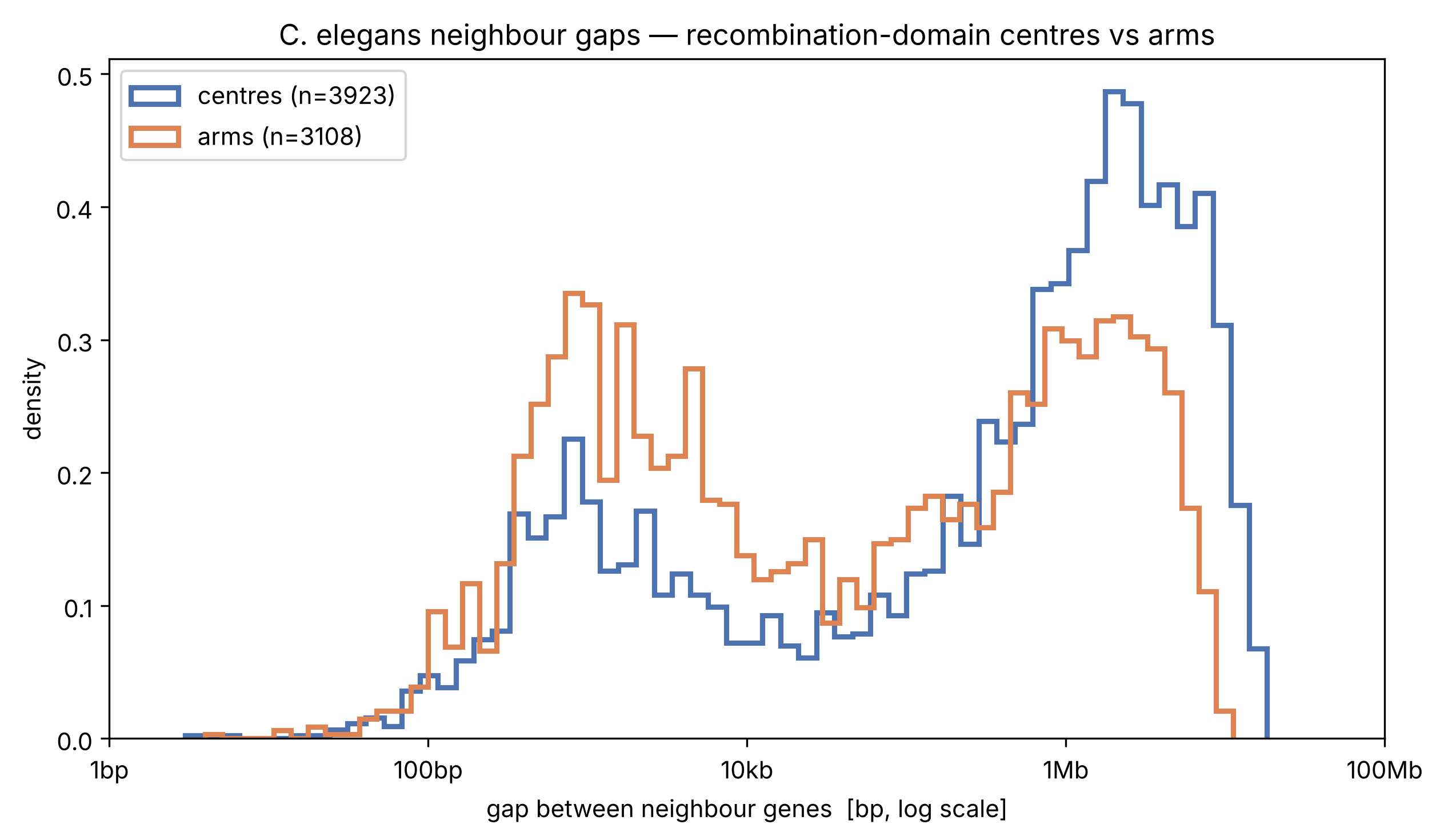}
\caption{Neighbour distances in \Ce{} split by recombination domain \cite{rockman}. The
high-recombination arms carry a larger short-range population ($\wshort = 0.50$) than the
gene-dense centres (0.32), and a shorter long-range one, since an arm offers 3.3--6.1~Mb of
room against 6.3--10.7~Mb for a centre.}
\label{fig:celegans}
\end{figure}

\paragraph{A within-genome test in \Ce.} \Ce{} chromosomes have gene-dense
centres with low recombination, flanked by arms with high recombination \cite{rockman}. Pairs on the arms are far more
often short-range ($\wshort = 0.50$, against 0.32 in the centres; Fig.~\ref{fig:celegans}),
consistent with the concentration of rapidly duplicating chemoreceptor families on the arms
\cite{robertson}. H3 predicts slightly longer short-range distances on the arms, where genes
are 25\% further apart (next-gene distance 954 vs 762~bp). The raw short-range distances
point that way (median 1,540 vs 1,212~bp, Mann--Whitney $p = 0.009$, a ratio of 1.27 against
a spacing ratio of 1.25), but the fitted modes (1,857 vs 1,936~bp) cannot resolve a contrast
this small. The test is consistent with H3 but could not have refuted it.

\section{Component weights: group size, then clade}
\label{sec:weights}

Raw short-range weights range from 0.26 (yeast, \At) to 0.71 (mouse), and human sits at
0.37. Three explanations are confounded: the group source, the size distribution of
groups, and clade. Human separates the first two, because it has two group tables on the
same coordinates and arms (Table~\ref{tab:weights}a). HGNC groups give $\wshort = 0.37$ and
InterPro families 0.62, but restricted to groups of 2--10 members they give 0.635 and
0.643. The group source makes no difference; the size distribution does. The short-range
mode is barely affected (22.9, 23.8, 18.8 and 23.3~kb), so the conclusions of
\S\ref{sec:h3} stand.

\begin{table}[!htbp]
\centering
\small
\caption{Short-range weight $\wshort$ after matching group size. (a)~Human, the same
coordinates and p/q arms, two group sources. (b)~All eight genomes, groups of 2--10 placed
members, with the range over four size windows (2--5, 2--10, 2--20, 3--10).}
\label{tab:weights}
\begin{tabular}{@{}llrrrr@{}}
\toprule
\multicolumn{6}{@{}l}{(a) Human} \\
Group table & & Mean size & Pairs & $\wshort$ & Short mode \\
\midrule
HGNC groups, all & & 16.3 & 20,554 & 0.372 & 22.9 kb \\
HGNC groups, 2--10 members & & 4.4 & 1,141 & 0.635 & 18.8 kb \\
InterPro Family, all & & 4.9 & 6,038 & 0.618 & 23.8 kb \\
InterPro Family, 2--10 members & & 3.2 & 1,723 & 0.643 & 23.3 kb \\
\midrule
\multicolumn{6}{@{}l}{(b) Groups of 2--10 members} \\
Species & Clade & Mean size & Pairs & $\wshort$ & Range over windows \\
\midrule
yeast       & non-vertebrate & 2.8 & 279   & 0.221 & 0.18--0.23 \\
\At         & non-vertebrate & 3.8 & 3,348 & 0.231 & 0.23--0.27 \\
\Ce         & non-vertebrate & 3.4 & 1,618 & 0.274 & 0.27--0.31 \\
fly         & non-vertebrate & 4.3 & 1,705 & 0.289 & 0.26--0.29 \\
maize       & non-vertebrate & 4.0 & 2,142 & 0.307 & 0.29--0.36 \\
zebrafish   & vertebrate     & 3.5 & 1,655 & 0.444 & 0.42--0.44 \\
mouse       & vertebrate     & 3.3 & 1,916 & 0.629 & 0.61--0.63 \\
human (InterPro) & vertebrate & 3.2 & 1,723 & 0.643 & 0.62--0.66 \\
\bottomrule
\end{tabular}
\end{table}

Matched for size, the genomes separate cleanly by clade (Table~\ref{tab:weights}b): the
five non-vertebrates have $\wshort = 0.22$--0.31 and the three vertebrates 0.44--0.64, with
no overlap under any of the four size windows (closest: maize 0.360 at 2--5 members against
zebrafish 0.416). The direction of the size bias depends on the genome. In human, large
groups dilute the tandem signal (0.37 raw, 0.64 matched); in \Ce, the largest groups are
the tandem-arrayed chemoreceptor families, so matching lowers the weight (0.40 raw, 0.27
matched). The three vertebrates are also three of the four Type~II genomes: a larger share
of tandem pairs, concentrated in particular families on particular chromosomes, is one way
per-chromosome mixtures come to differ.

\section{Discussion}

The study began with a picture. Drawing every gene of eight genomes on ideograms, and the
members of any fly gene group on a banded ideogram in an interactive application, made the
spatial distribution of gene families something one can browse, group by group. Two
impressions from browsing --- members that look evenly spaced, and tight clusters next to
megabase-scale spreads --- were then measured.

Evenly spaced members turned out to be the weaker of the two. Sets can be strikingly
symmetric --- in the RAB GTPases two spacings of about 5~Mb differ by only 634~bp, about
0.01\% --- but an arm-matched random control
produces about 85\% as many, and a minimum-distance filter that hides tandem clusters turns
a modest enrichment into an apparent depletion. The few symmetric sets that chance does not
produce are tandem-scale ones such as bcd--Dfd--Scr in the Antennapedia complex, and these
are better described by their short distances than by their symmetry. Visual patterns on an
ideogram need the same random control as any other statistic.

The distances between neighbour genes have the same structure across eight eukaryotic
genomes, three kingdoms (animals, plants and fungi) and a 250-fold range of genome size: a short-range population and a
long-range population. The random control in fly and the per-unit fits in the other genomes
assign them different meanings. The long-range population is what randomly placed genes
produce: it is reproduced by the control in fly, nearly exponential where it dominates (yeast), and
its scale follows the length of the chromosome or arm. The short-range population is what
random placement cannot produce, and it is the signature of family members that lie close
together, most plausibly as products of tandem duplication \cite{zhang,shoja,rizzon}.

Fitted chromosome by chromosome, the genomes fall into two types. In Type~I genomes (fly,
maize, \At, \Ce) the per-unit mixtures nearly coincide, because their units are of similar
length and, apart from the chemoreceptor-rich arms of \Ce, their tandem families are spread
fairly evenly across them. In Type~II genomes (human, mouse,
zebrafish, yeast) they vary, for three separable reasons: a wide range of unit lengths moves
the long-range peak (human, mouse); tandem-expanded families concentrated on particular
chromosomes move the short-range weight (human \texttt{19}, zebrafish \texttt{4}); and too
few pairs per chromosome make the short-range fit noisy (yeast). The typology is therefore
descriptive, not a single biological property, and its interest lies in which of these
reasons applies. The division nonetheless echoes the companion comparison of the same eight
genomes \cite{wu2026}, in which gene-family and protein-function composition was more uniform
along the chromosomes of the plants and the fly than along those of the vertebrates; there,
\Ce{}, the borderline Type~I genome here, was the least uniform of all.

The scale of the short-range population follows the spacing of genes in each genome. This is not
obvious in advance. One might expect it to follow genome size, since larger genomes have
more DNA between everything, or gene size, since a tandem copy must clear the gene it
copies. Maize and zebrafish separate these alternatives because they break the usual
correlation between genome size, gene size and gene spacing in opposite directions. Maize
packs plant-sized genes into mammal-sized spacing, its large genome consisting mostly of
retrotransposons between genes \cite{schnable}. Zebrafish has the longest genes in the set
and less intergenic DNA than the mammals. In both, the short-range mode follows the
spacing. A natural reading is that a tandem copy, once it has drifted away from its
original by the insertion of other genes, other sequences or rearrangements, ends up a
couple of gene spacings away, so the clustering scale is measured in genes, not in base
pairs.

The weights of the two populations are a different matter. Raw weights mostly reflect how
large the groups in each table are, a point that matters for any comparison of
clustering between genomes annotated by different resources. Once group size is matched, the
three vertebrates carry about twice the share of short-range pairs found in the five
non-vertebrates. Why is open. The three vertebrates share ancestry, so this is one contrast,
not three independent observations, and it could reflect lineage-specific expansions of
large tandem families (olfactory receptors, zinc-finger genes) \cite{zhangfirestein,huntley}.

\section{Caveats}

\begin{itemize}
\item \emph{Group definitions differ.} Fly and human use curated tables; the other six use
InterPro protein families, which miss non-coding families and can split curated groups.
Two-source comparisons in human and \Ce, and the \emph{Family}~+~\emph{Domain} check, show
that the modes are stable to within 3--40\% in all genomes except maize (a factor of two).
\item \emph{Units differ.} Fly and human pairs are confined to arms; elsewhere a pair may span
a centromere, which inflates long-range distances. Long-range modes are compared only within
a unit convention. Splitting \At{} and maize chromosomes at their centromeres is an obvious
refinement.
\item \emph{The Type~I/Type~II division is descriptive.} Its thresholds were set at the break
between the observed genomes, \Ce{} lies close to it, and the per-unit SDs depend on the
number of pairs per unit (yeast). It summarises the per-unit fits; it is not a test.
\item \emph{The application covers fly only.} Symmetric sets were counted and controlled in
fly; the same functions run on the other genomes, but no interactive view or random control
was built for them.
\item \emph{Annotation density differs.} Mammalian annotations contain many nested and
antisense genes (29--34\% of adjacent pairs overlap). Dropping them biases the mammalian
next-gene distance downward; correcting it would widen, not narrow, the spread in
Table~\ref{tab:h3}.
\item \emph{Few genomes.} H3 rests on eight points, with no genomes at 0.9--6.9~kb
and 6.9--14~kb spacing. A rank correlation of $\rho = 0.98$ at eight genomes (one adjacent
swap) has an exact two-sided permutation $p$ of about $4\times10^{-4}$, and the yeast and fly
short-range modes (1,768 and 1,769~bp) are effectively tied; it is a pattern worth testing on more genomes, not a scaling law.
\item \emph{The random control was run only in fly.} The reading of the long-range component
as random placement in other genomes rests on its scaling with unit length and, in yeast,
on a back-of-envelope estimate. Running the control in each genome is the most direct
next step.
\item \emph{Fly gene biotypes} in the arm statistics of the Appendix were inferred from
FlyBase symbol prefixes, not from the GFF3 biotype field.
\item \emph{Human coordinates} were re-downloaded after an initial BioMart download silently
dropped chromosomes 1 and 2; all numbers here are from the corrected data.
\end{itemize}

\section*{Data and code availability}
\label{sec:data}

\begin{sloppypar}
All analyses use public releases: FlyBase FB2026\_01 gene list and gene-group report (dm6)
\cite{flybase}; HGNC complete set \cite{hgnc}; Ensembl BioMart (GRCh38, GRCm39, GRCz11,
R64-1-1, WBcel235) \cite{ensembl}; Ensembl Plants GFF3 (TAIR10, Zm-B73-REFERENCE-NAM-5.0)
\cite{ensemblgenomes}; UniProt reference proteomes UP000006548 (\At) and UP000001940 (\Ce)
\cite{uniprot}; the InterPro entry list \cite{interpro}; WormBase gene classes
\cite{wormbase}; and the dm6 cytogenetic band map distributed with \texttt{ideogram.js}
\cite{ideogram}. Data were downloaded between May and September 2026. No new sequence data
were generated. All figures are produced by the Jupyter notebooks \texttt{01\_ideogram},
\texttt{02\_gene\_markers}, \texttt{03\_ideogram\_app}, \texttt{04\_chromosome\_arm\_stats},
\texttt{05\_random\_control\_gene\_distances}, \texttt{06\_fly\_neighbour\_gene\_distances},
\texttt{07}--\texttt{12\_<species>\_neighbour\_gene\_distances} (human, mouse, zebrafish,
yeast, \At, maize), \texttt{13\_gene\_class1\_ideogram\_all\_species} and
\texttt{14\_celegans\_neighbour\_gene\_distances}, using the \texttt{fly\_genome.io},
\texttt{fly\_genome.viz} and \texttt{fly\_genome.neighbour} modules. The interactive
application is notebook \texttt{03\_ideogram\_app}, served with \texttt{panel serve}; its
screenshots (Figs.~\ref{fig:app} and \ref{fig:appsym}) were taken at its default settings,
and the ideograms of Figs.~\ref{fig:appsym} and \ref{fig:topsym} were redrawn with the same
plotting functions so that gene names are shown for every group. The source code is
available from the author upon reasonable request.
\end{sloppypar}

\section*{Acknowledgements}

This work used no funding and no computing resources beyond a laptop. Analyses were
carried out in Python with \texttt{numpy} \cite{numpy}, \texttt{pandas} \cite{pandas}, \texttt{scipy}
\cite{scipy}, \texttt{scikit-learn} \cite{sklearn}, \texttt{matplotlib} \cite{matplotlib}
and \texttt{panel} \cite{panel}.

The author would like to sincerely thank Anthropic's Claude Opus~5.5 for its incredibly
intelligent aid in the baking of this manuscript.

\appendix
\renewcommand{\thefigure}{A\arabic{figure}}
\setcounter{figure}{0}
\section*{Appendix: supplementary figures}
\addcontentsline{toc}{section}{Appendix}

\begin{figure}[!htbp]
\centering
\includegraphics[width=\textwidth]{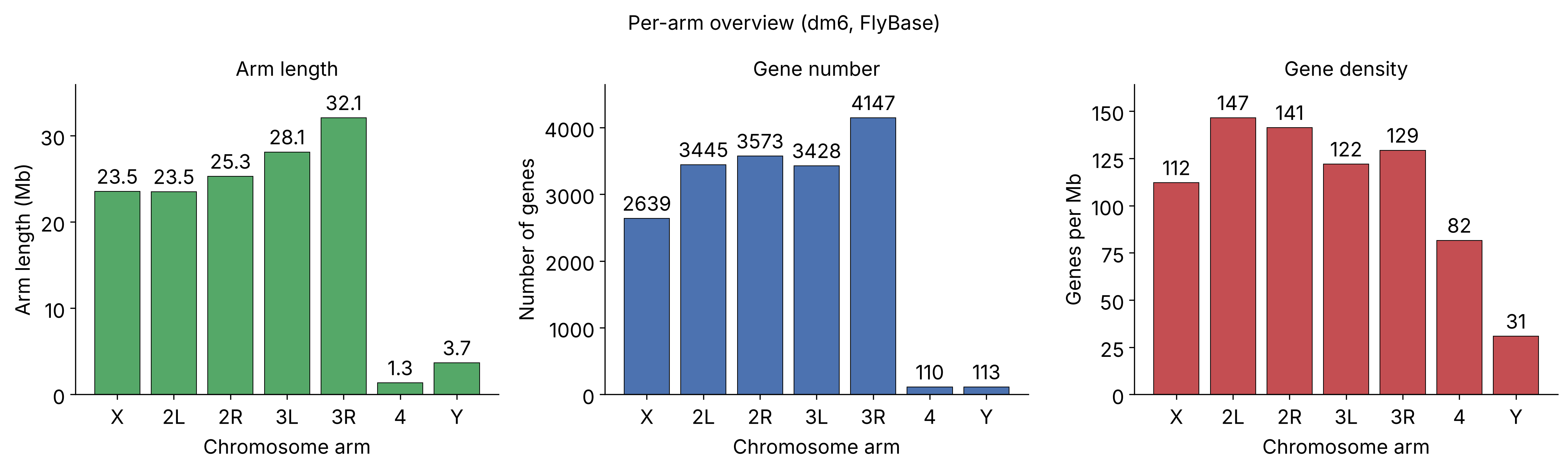}\\[4pt]
\includegraphics[width=\textwidth]{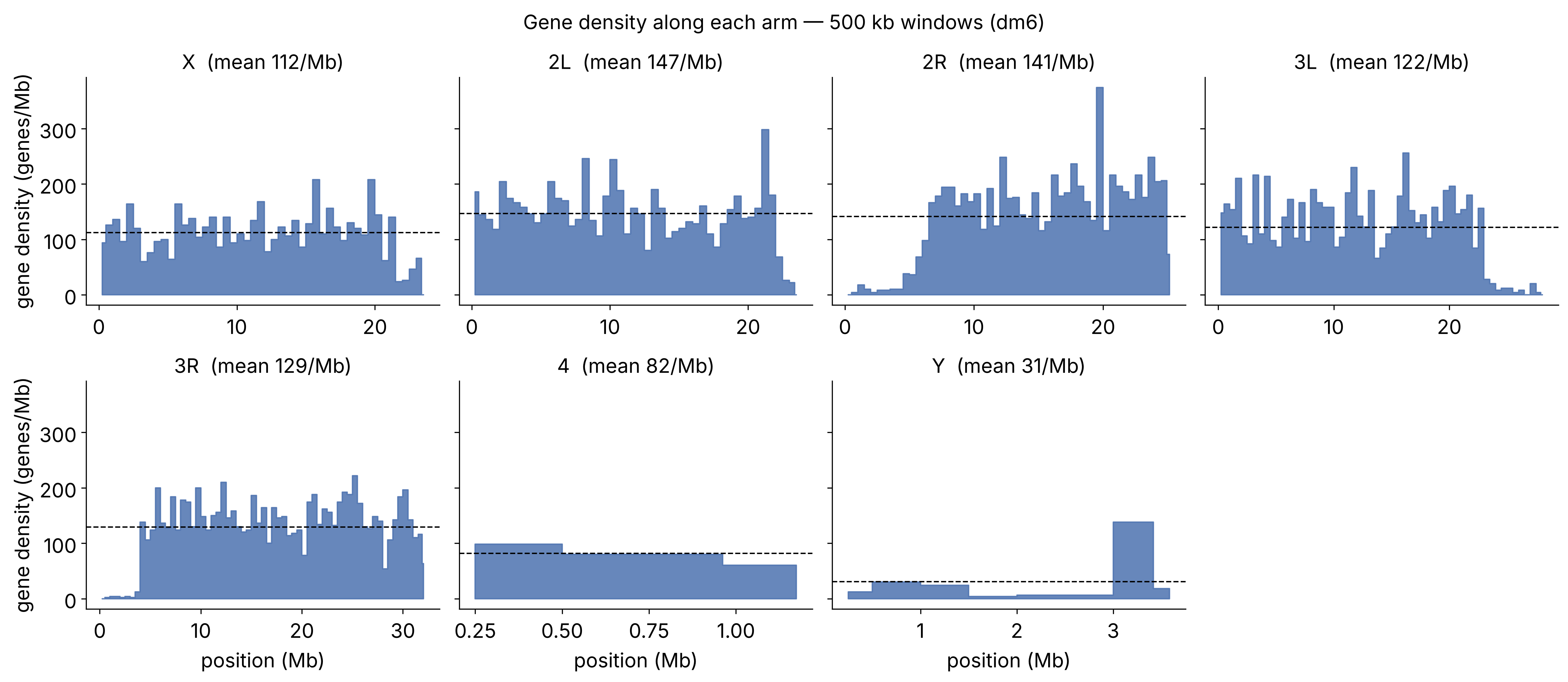}
\caption{Fly chromosome arms (dm6, FlyBase). Top: arm length, gene number and gene density.
Bottom: gene density along each arm in 500~kb windows (dashed: arm mean). Density falls
towards the centromere-proximal end of each arm: near coordinate~0 on \texttt{2R} and
\texttt{3R}, and at the high-coordinate end on \texttt{X}, \texttt{2L} and \texttt{3L}.}
\label{fig:armstats}
\end{figure}

\begin{figure}[!htbp]
\centering
\includegraphics[width=\textwidth]{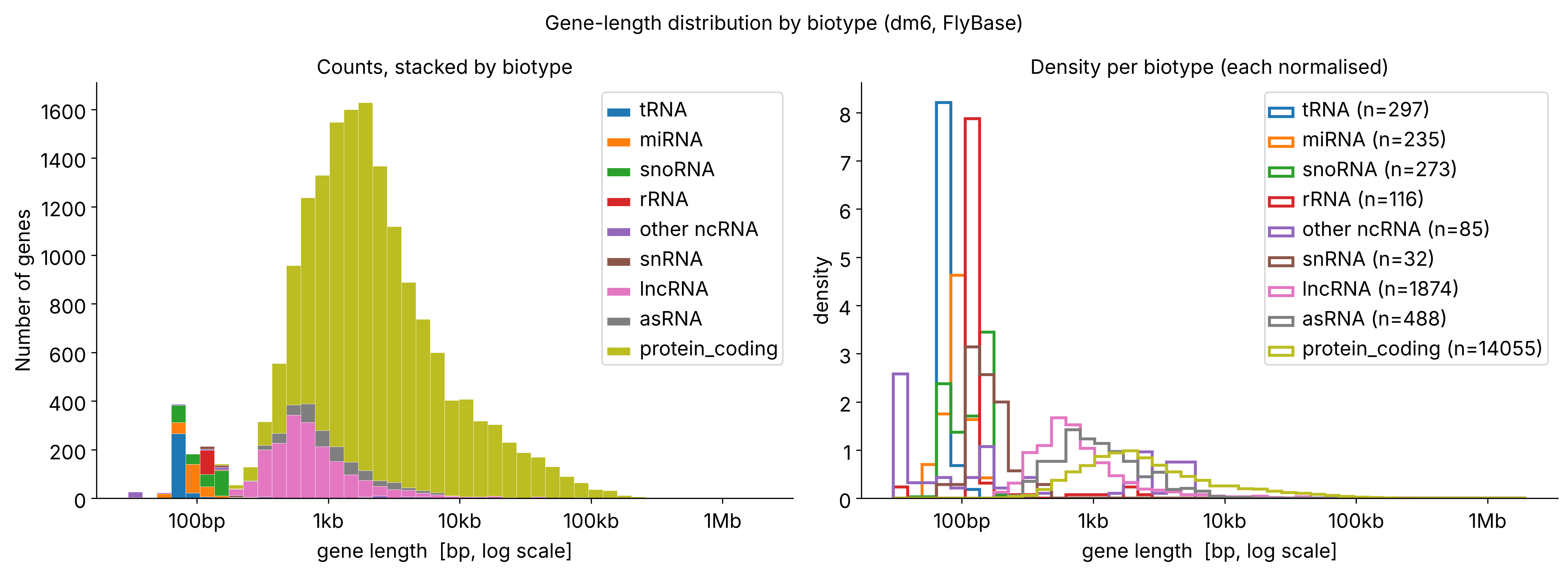}
\caption{Fly gene length by biotype. Left: counts, stacked; right: each biotype normalised.
The peak near 100~bp is formed by tRNA (median 73~bp), miRNA (95), snoRNA (131), rRNA (135)
and snRNA (154~bp); lncRNA (658~bp), antisense RNA (1.0~kb) and protein-coding genes
(2.1~kb, $n = 14{,}055$) follow. Biotypes were inferred from FlyBase symbol prefixes, so
counts differ slightly from the FlyBase gene classes of Fig.~\ref{fig:ideogram}b (e.g.\
14,055 against 13,945 protein-coding genes).}
\label{fig:biotype}
\end{figure}

\begin{figure}[!htbp]
\centering
\includegraphics[width=\textwidth]{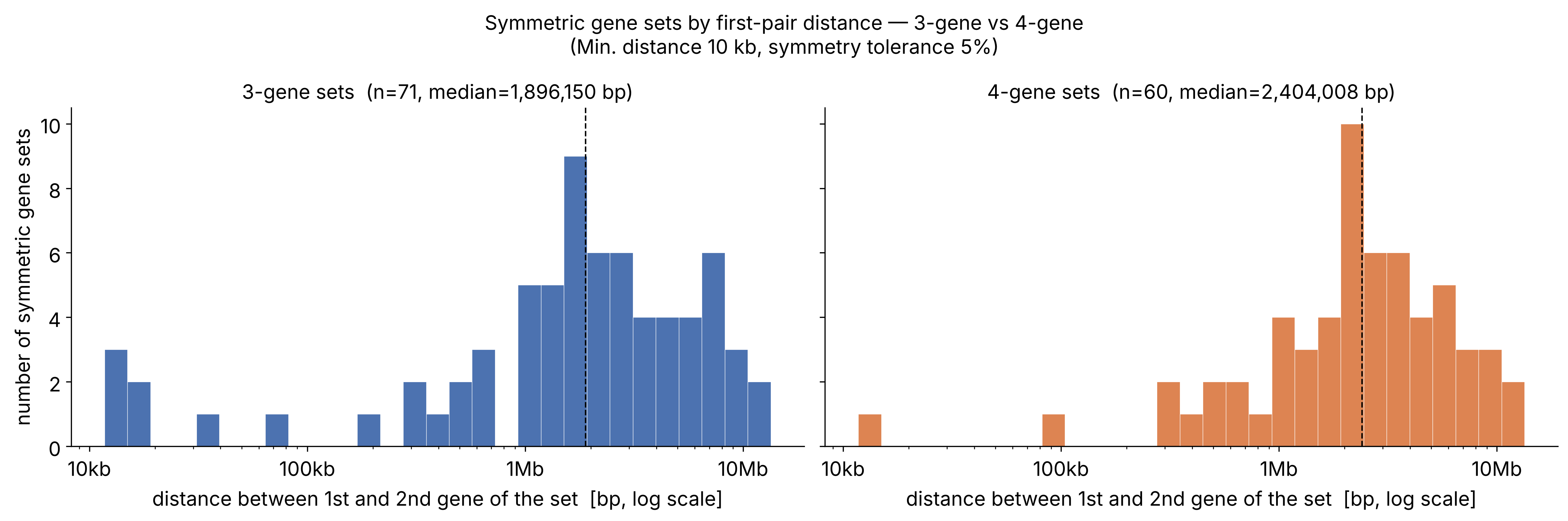}
\caption{First spacing of the 131 fly symmetric sets that pass a 10~kb minimum distance and
a 5\% tolerance: 71 three-gene sets (left) and 60 four-gene sets (right); dashed, median
(about 1.9 and 2.4~Mb).}
\label{fig:symfirst}
\end{figure}

\begin{figure}[!htbp]
\centering
\includegraphics[width=\textwidth]{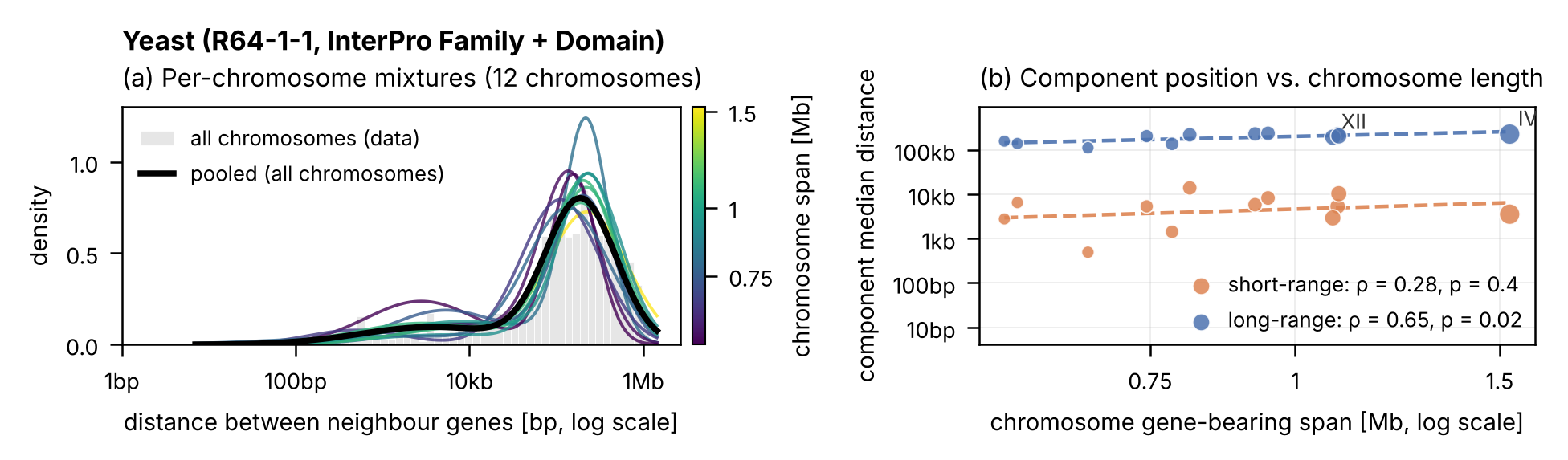}
\caption{Yeast with the looser InterPro \emph{Family}~+~\emph{Domain} groups (12 chromosomes
with at least 50 positive distances), drawn as the rows of Fig.~\ref{fig:typeII}:
(a)~per-chromosome mixtures; (b)~component medians against chromosome span (long-range
$\rho = 0.65$, $p = 0.02$).}
\label{fig:yeastfd}
\end{figure}

\FloatBarrier

\end{document}